\documentclass[aps,prb,onecolumn,superscriptaddress,floatfix,notitlepage,11pt,tightenlines]{revtex4-1}
\makeatletter
\AtBeginDocument{\let\LS@rot\@undefined}
\makeatother
\usepackage{color}
\usepackage{dcolumn} % decimal align in tables
\usepackage{bm} % bold math
\usepackage{graphicx}
\usepackage{multirow} % for table cells to span rows
\usepackage{pifont} % for checkmarks
\usepackage{epsfig}
\usepackage{verbatim}
\usepackage{amsfonts}
\usepackage{amsmath} % matrix
\usepackage{amssymb}
\usepackage{mathtools}
\usepackage{subcaption}
\usepackage{braket}
\usepackage{float}
\usepackage{pdfpages}
\usepackage{pdflscape}
\usepackage[version=4]{mhchem}
\usepackage{array}
\usepackage[columnwise]{lineno}
\usepackage{mathrsfs}
\usepackage{epigraph}
\usepackage[mathcal]{euscript}
\usepackage{textcomp}
\usepackage{threeparttable}
\usepackage{enumitem}
\usepackage{caption}
\usepackage{url}
\usepackage{comment}
\usepackage{float}
\usepackage{wrapfig}
\usepackage{xcolor,listings} % for color C/C++, python etc. codes
\usepackage[nolist]{acronym}
\usepackage[titletoc,toc,title]{appendix}
\usepackage[bf,raggedright]{titlesec}
\usepackage{textcomp}
\usepackage[left=1.0in,right=1.0in,top=1.0in,bottom=1.0in]{geometry}
\usepackage{indentfirst}	%indenting the first titles
\newcommand{\etal}{\emph{et~al.}}

\newcommand{\si}{\text{Supporting Information}}

\newcommand{\al}{\alpha}

\newcommand{\la}{\lambda}

\newcommand{\ze}{\zeta}
\newcommand{\om}{\omega}

\newcommand{\sig}{\sigma}

\newcommand{\DE}{\Delta E}

\newcommand{\Req}{R_{\text{eq}}}

\newcommand{\IEKTm}{I_{1,m}^{\text{EKT}}}
\newcommand{\IEKTk}{I_{k}^{\text{EKT}}}
\newcommand{\IEKTCAS}{I_{1,\text{CAS-CI}}^\text{EKT}}
\newcommand{\IEKTFCI}{I_{1,\text{FCI}}^\text{EKT}}
\newcommand{\IHCIk}{I_k^\text{HCI}}
\newcommand{\TAUCAS}{\tau_\text{CAS-CI}}
\newcommand{\TAUFCI}{\tau_\text{FCI}}

\newcommand{\EKTCASCI}{\DE_{1,\text{CAS-CI}}}
\newcommand{\EKTFCI}{\DE_{1,\text{FCI}}}
\newcommand{\EKTm}{\DE_{1,m}}

\newcommand{\Etot}{E_\text{tot}}

\newcommand{\MAE}{\overline{\text{MAE}}}
\newcommand{\MAEe}{\text{MAE/e}}
\newcommand{\AKEEe}{\text{AKEE/e}}
\newcommand{\MAKEE}{\overline{\text{MAKEE}}}

\newcommand{\NIAD}{\overline{\text{NIAD}}}

\newcommand{\Eh}{E_\text{h}}

\newcommand{\wn}{\text{cm}^{-1}}
\newcommand{\bD}{\mathbf{D}}

\newcommand{\bK}{\mathbf{K}}

\newcommand{\bc}{\mathbf{c}}

\newcommand{\br}{\mathbf{r}}

\newcommand{\gamess}{\textsc{GAMESS--US}}

\setenumerate[1]{label=\thesection.\arabic*.}
\newcommand{\mr}{\multirow}

\begin{document}
%---------------------------------------------------
%                Equations skip control
%---------------------------------------------------
%\setlength{\abovedisplayskip}{4pt plus 1.5pt minus 2pt}
%\setlength{\belowdisplayskip}{4pt plus 1.5pt minus 2pt}
%\setlength{\abovedisplayskip}{4pt plus .7pt minus 2pt}
%\setlength{\belowdisplayskip}{4pt plus .7pt minus 2pt}
%---------------------------------------------------

%=================================
% Affiliations
%---------------------------------

\author{Reza Hemmati$^{1,2,\dagger,*}$, Mohammad Mostafanejad$^{1,2,\ddagger,*}$, and J. V. Ortiz$^{3,\star}$ \\[3pt]
\small $^{1}$\textit{Department of Chemistry, Virginia Tech, Blacksburg, Virginia 24061, USA} \\
\small $^{2}$\textit{Molecular Sciences Software Institute, Blacksburg, Virginia 24060, USA}\\
\small $^{3}$\textit{Department of Chemistry and Biochemistry, Auburn University, \\ 
Auburn, Alabama, 36849-5312, USA}\\
\small \textit{E-mails: $^{\dagger}$rhemmati@vt.edu, $^{\ddagger}$smostafanejad@vt.edu, $^{\star}$ortiz@auburn.edu}, \\ 
\small $^{*}$Equal contributions \\[3pt]}

% \author{Reza Hemmati}
% \email{rhemmati@vt.edu}
% \thanks{Equal contribution}
% \author{Mohammad Mostafanejad}
% \thanks{Equal contribution}
% \email{smostafanejad@vt.edu}
% \affiliation{Department of Chemistry, Virginia Tech, Blacksburg, Virginia 24061, USA}
% \affiliation{Molecular Sciences Software Institute, Blacksburg, Virginia 24060, USA}
% %---------------------------------
% \author{Joseph Vincent Ortiz}
% \email{ortiz@auburn.edu}
% \affiliation{Department of Chemistry and Biochemistry, Auburn University, Auburn, Alabama, 36849-5312, USA}
%=================================

%=================================
\title{Numerical Analysis of the Complete Active-Space Extended Koopmans's Theorem}
%=================================

%=================================
% Abstract
%---------------------------------
\acresetall	% reset all acronyms
%---------------------------------
\begin{abstract}
    We investigate the numerical accuracy of the \ac{EKT} in reproducing the
    \ac{FCI}\acused{CI} and \ac{CASCI} \acp{IE} of atomic and molecular systems
    calculated as the difference between the energies of $N$ and $(N-1$) electron
    states. In particular, we study the convergence of the \ac{EKT} \acp{IE} to
    their exact values as the basis set and the active space sizes vary. We find
    that the first \ac{FCI} \ac{EKT} \acp{IE} approach their exact counterparts as
    the basis set size increases. However, increasing the basis set or the active
    space sizes do not always lead to more accurate \ac{CASCI} \ac{EKT} \acp{IE}.
    Our investigation supports the Davidson \etal's\ observation [E. R. Davidson,
    \etal, J. Chem. Phys. 155, 051102 (2021)] that the \ac{FCI} \ac{EKT} \acp{IE}
    can be systematically improved with arbitrary numerical accuracy by
    supplementing the basis set with diffuse functions of appropriate symmetry which
    allow the detached electron to travel far away from the reference system. By
    changing the exponent and the center of the diffuse functions, our results
    delineate a complex pattern for the \ac{CASCI} \ac{EKT} \ac{IE} of \ce{LiH}
    which can be important for the spectroscopic studies of small molecules.
\end{abstract}
%=================================
\maketitle

%=================================
% Main Text
%---------------------------------
\acresetall	% reset all acronyms
%---------------------------------
%=============================================
\section{Introduction}\label{SEC:INTRODUCTION}
%---------------------------------------------
The \ac{IE} and \ac{EA} of atoms and molecules can be computed from the total
energies of the exact $N$-electron reference wavefunction and its ($N\! \pm\!
  1$)-electron counterpart. However, finding the exact solution to the many-body
Schr\"{o}dinger equation is infeasible for all but few
systems.\cite{Helgaker:2000:BOOK} \ac{KT}\cite{Koopmans:1934:104} offers a
computationally affordable model for approximating the \ac{IE} and \ac{EA} of
atoms and molecules based on the \ac{HF} description of the system. The
simplicity of the \ac{KT}, however, comes with major deficiencies: The Coulomb
electron correlation effects\cite{Avella:2012:BOOK} are not captured by the
model and the electronic relaxation effects, driven by the removal or the
addition of an electron, are ignored.\cite{Smith:1975:113}

In order to mitigate the aforementioned problems in \ac{KT}, the
\ac{EKT}\cite{Day:1974:501,Smith:1974:511,Morrell:1975:549,Smith:1975:113}
employs a post-\ac{HF} reference wavefunction to incorporate the electron
correlation and accommodates the relaxation effects, variationally. These
improvements have contributed to the popularity of \ac{EKT} which has been
extensively benchmarked against highly accurate quantum chemistry methods and
experimental data.\cite{Day:1975:115,Ellenbogen:1977:4795,Morrison:1992:3718,
  Sundholm:1993:3999,Cioslowski:1997:6804,Bozkaya:2014:2041,Bozkaya:2018:4375,
  Pomogaev:2021:9963,Lee:2021:3372,Sabatino:2021:746735,Ortiz:2022:109} Despite
its successful performance in a wide range of applications, the exactness of the
\ac{EKT} method, especially for predicting the lowest \ac{IE} of many-electron
systems, has been a subject of intense scrutiny and
debate.\cite{Pickup:1975:422,Morrell:1975:549,Andersen:1977:1067,
  Katriel:1980:4403,Pickup:1988:69,Morrison:1993:6221,Sundholm:1993:6222,
  Morrison:1994:649,Ernzerhof:2009:793,Davidson:2021:051102} Based on the
asymptotic behavior of the electron density of atoms and molecules, from which
an electron is removed, Morell, Par and Levy\cite{Morrell:1975:549} established
the exactness of the \ac{EKT} for the lowest \ac{IE}. This finding was further
corroborated by several authors.\cite{Ahlrichs:1976:2706,Levy:1976:2707,
  Katriel:1980:4403} However, independent studies on the perturbative expansions
of the exact \acp{IE} revealed that the \ac{EKT} \acp{IE} cannot be exact due to
the missing second-order terms.\cite{Pickup:1975:422,Pickup:1988:69} The
conclusions drawn from these studies have been both
supported\cite{Andersen:1977:1067} and refuted\cite{Olsen:1998:282} by
subsequent investigations.

A different view point is offered by Ernzerhof\cite{Ernzerhof:2009:793} who
argued that the question of the exactness of the \ac{EKT} is not well-posed as
the lowest eigenvalue of the \ac{EKT} generalized eigenvalue equation may not
exist. In particular, for all finite distances of the $N$th electron from its
parent ($N\!-\!1$)-electron system during the electron removal process, none of
the \ac{EKT} eigenvalues become equal to the exact lowest \ac{IE} of the
system.\cite{Ernzerhof:2009:793} However, if the system possesses a diffuse
basis function which allows the electron to travel very far from the
($N\!-\!1$)-electron system, the \ac{EKT} eigenvalues can get arbitrarily close
to the exact \ac{IE} as the $N$th electron goes to
infinity.\cite{Ernzerhof:2009:793} This conjecture has been further studied by
Davidson \etal\cite{Davidson:2021:051102} who focused on the first \ac{EKT}
\ac{IE} of the beryllium atom.

In this effort, we aim to further investigate Ernzerhof's
conjecture\cite{Ernzerhof:2009:793} using the \ac{CASCI} and \ac{FCI} reference
wavefucntions and a comprehensive set of correlation-consistent basis
functions.\cite{Dunning:1989:1007,Woon:1994:2975,Prascher:2011:69,Peterson:2002:10548,
  Kendall:1992:6796} Furthermore, we expand the scope of Davidson's
work\cite{Davidson:2021:051102} to a wider range of atomic and molecular systems
and study the impact of various active space and basis set sizes on the
predicted \ac{EKT} \acp{IE}. Finally, we generate a three-dimensional surface
for the first \ac{EKT} ionization energy of the \ce{LiH} molecule and
discuss the importance of the diffuse basis function on the calculated results.

The manuscript is organized as follows. In Sec.~\ref{SEC:THEORY}, we overview
the important theoretical aspects of the \ac{EKT} and the nuances surrounding
its exactness. We also introduce an error metric, based on an energy bound
inequality involving the first \ac{EKT} \ac{IE}, in order to be able to quantify
the discrepancy between the \ac{EKT} and the conventional quantum mechanical
method of calculating the \ac{IE}. Technical details pertinent to the
calculations performed in this study are presented in
Sec.~\ref{SEC:COMPDETAILS}. In Sec.~\ref{SEC:RESULTS} we discuss the results of
our investigation and provide a detailed analysis of the first \ac{EKT}
ionization energies of the \ce{Be} atom, \ce{H-} ion, \ce{H2}, \ce{Li2} and the
\ce{LiH} molecule. We conclude our study in Sec.~\ref{SEC:CONCLUSION} with a
summary of our findings and the implications of our results.
%=============================================

%=============================================
\section{Theory}\label{SEC:THEORY}
%---------------------------------------------
Let $\ket{\Psi^N}$ be the normalized ground-state wavefunction of an
$N$-electron system. The central assumption in \ac{EKT} is that the wavefunction
of the $k$th state of the $N-1$ electron ionic system, $\ket{\Psi_k^{N-1}}$, can
be approximated as a linear combination of hole-state \acp{CSF},
$\ket{\tilde{\Psi}^{N-1}_k}$, as\cite{Day:1974:501,Smith:1974:511,
  Morrell:1975:549,Smith:1975:113}
\begin{equation}\label{EQ:PSITILDE}
  \ket{\tilde{\Psi}^{N-1}_k} = \hat{A}_k \ket{\Psi^N}.
\end{equation}
Here, the electron removal operator $\hat{A}_k$ is defined as
\begin{equation}\label{EQ:AK}
  \hat{A}_k = \sum_{j} c_{jk} \hat{a}_j,
\end{equation}
in which, the operator $\hat{a}_j$ annihilates an electron from the $j$th
spin-orbital, $\phi_j$, of the reference $N$-electron wavefunction,
$\ket{\Psi^N}$. Adopting Eq.~\ref{EQ:PSITILDE}, the \ac{HCI} ionization
energy of the $k$th state of the $N-1$ electron system can be written
as
\begin{equation}
  \begin{aligned}\label{EQ:IEKT1}
    \IHCIk & = \tilde{E}_k^{N-1} - E^N                                              \\
           & = \frac{\braket{\Psi^N | \hat{A}_k^{\dag} \hat{H} \hat{A}_k | \Psi^N}}
    {\braket{\Psi^N_k | \hat{A}_k^{\dag} \hat{A}_k |\Psi^N_k}}
    - \braket{\Psi^N | \hat{H} | \Psi^N} \qquad \text{where} \qquad k = 1, 2, 3, \ldots .
  \end{aligned}
\end{equation}
The subscript $k=1$ corresponds to the ground-state of the ionic system and the
superscript ``$\dag$'' stands for the self-adjoint operation. Solving
Eq.~\ref{EQ:IEKT1} involves dealing with 3-particle \acp{RDM} which are
expensive to compute and inconvenient to handle. The \ac{EKT} model circumvents
this hurdle by assuming that the reference wavefunction $\ket{\Psi^N}$ is an
eigenfunction of the exact or model Hamiltonian. As such, the \ac{EKT}
ionization energy can be expressed as
\begin{equation}\label{EQ:IEKT2}
  \IEKTk = \frac{\braket{\Psi^N | \hat{A}_k^{\dag} [ \hat{H}, \hat{A}_k ] | \Psi^N}}
  {\braket{\Psi^N_k | \hat{A}_k^{\dag} \hat{A}_k |\Psi^N_k}}
\end{equation}
where the commutator $[\cdot,\cdot]$ is defined as $[\hat{A},\hat{B}] =
  \hat{A}\hat{B} - \hat{B}\hat{A}$.\cite{Szabo:1996:BOOK} The aforementioned
assumption holds if the reference wavefunction is exact or corresponds to, for
example, the \ac{HF}, \ac{CASCI}, or the \ac{FCI} methods. It is important to
note that, in general, the \acused{CI}\ac{CI} reference wavefunctions may not be
the eigenfunctions of the Hamiltonian, represented in a complete or truncated
basis set. In such cases, the non-Hermitian numerator of Eq.~\ref{EQ:IEKT2}
should be replaced by its Hermitian counterpart,

\begin{equation}\label{EQ:IEKT2H}
  \frac{1}{2} \braket{\Psi^N | \hat{A}_k^{\dag} [ \hat{H}, \hat{A}_k ]
  + [ \hat{A}_k^{\dag}, \hat{H} ] \hat{A}_k | \Psi^N}.
\end{equation}

Substituting Eq.~\ref{EQ:AK} into Eq.~\ref{EQ:IEKT2} and calculating the
variations of the \ac{EKT} \ac{IE} with respect to changes in the
electron removal expansion coefficients, $c_{jk}$, allow us to obtain the
\ac{EKT} generalized eigenvalue equation as
\begin{equation}\label{EQ:IEKT3}
  \bK \bc_k = \IEKTk \bD \bc_k.
\end{equation}

Here, $\bc_k$ stands for the $k$th column eigenvector of the generalized
eigenvalue equation, $\bK$ is the Koopmans's matrix, and $\bD$ denotes the
\ac{1-RDM}. The elements of the Koopmans's matrix and \ac{1-RDM} are defined as
\begin{equation}\label{EQ:KIJ}
  K_{ij} = \braket{\Psi^N | \hat{a}_i^{\dag} [\hat{H}, \hat{a}_j] | \Psi^N}
\end{equation}
and
\begin{equation}\label{EQ:DIJ}
  D_{ij} = \braket{\Psi^N | \hat{a}_i^{\dag} \hat{a}_j | \Psi^N},
\end{equation}
respectively. In solving the generalized eigenvalue equation \ref{EQ:IEKT3},
numerical difficulties may arise if symmetric
orthonormalization\cite{Szabo:1996:BOOK} is adopted and the natrual orbital
occupation numbers become very small.\cite{Morrison:1992:1004} Previous
experiments with alternative techniques such as canonical
orthonormalization\cite{Szabo:1996:BOOK} have negatively impacted the accuracy
of the predicted \ac{EKT} ionization energies due to the elimination
of the natural orbitals with the smallest occupation
numbers.\cite{Morrison:1992:1004,Morrison:1994:309,Davidson:2021:051102}
Regarding Eq.~\ref{EQ:IEKT3}, it is also important to note that in general, the
first \ac{EKT} ionization energy (or the lowest eigenvalue of the
Koopmans's matrix) is not exact or equal to the difference between the
corresponding \ac{FCI} total energies of the $N$- and ($N\!-\!1$)-electron
states.\cite{Ernzerhof:2009:793,Davidson:2021:051102} As a result, for the
first \ac{FCI} and \ac{CASCI} \ac{EKT} ionization energies, one can
write
\begin{equation}\label{EQ:INEQUALITY}
  \IEKTm \ge \EKTm,
\end{equation}
where
\begin{equation}\label{EQ:DEM}
  \EKTm = E_{1,m}^{N-1} - E_m^N.
\end{equation}
Here, $E_m^N$ and $E_{1,m}^{N-1}$ denote the total energies of the ground-state
of $N$- and ($N\!-\!1$)-electron systems, respectively and $m$ stands for
either \ac{FCI} or \ac{CASCI}. For two-electron systems ($N = 2$), the equality
holds for the first \ac{EKT} ionization energies of both \ac{FCI} and \ac{CASCI}
wavefucntions. Furthermore, the equality remains valid if the active space
pertinent to the \ac{CASCI} reference wavefunction consists of distributing $N$
electrons in $N/2$ active orbitals or CAS($N,\tfrac{N}{2}$) which corresponds to
a closed-shell system.\cite{Davidson:2021:051102} In order to quantify the
discrepancy between the calculated \ac{EKT} ionization energy,
$\IEKTm$, and $\EKTm$, we define our error metric, $\tau_m$, as
\begin{equation}\label{EQ:TAU}
  \tau_m = \IEKTm - \EKTm,
\end{equation}
which according to Eq.~\ref{EQ:INEQUALITY} must be non-negative.
%=============================================

%=============================================
\section{Computational Details}\label{SEC:COMPDETAILS}
%---------------------------------------------
%=============================================
All calculations are performed using \gamess\ quantum chemistry program package
version R2 (30 SEP 2022).\cite{Barca:2020:154102} We have modified the source
code to allow for the tolerance of natural orbital occupation numbers as small
as $10^{-10}$ when solving the \ac{EKT} equation (Eq.~\ref{EQ:IEKT3}). The
energy convergence threshold is set to $10^{-7} \Eh$ for all \ac{FCI} and
\ac{CASCI} calculations in Sec.~\ref{SUBSEC:ATOMIONIZATION}. Dunning's
(augmented-)correlation consistent basis sets, (aug-)cc-pVXZ and (aug-)cc-pCVXZ
with X = D, T, Q, and 5, are employed for running the aforementioned
calculations.\cite{Dunning:1989:1007, Woon:1994:2975,Prascher:2011:69} We used
Dask\cite{DASK:2016:URL} for handling the thousands of \ac{CASCI} computations
required for generating the three-dimensional \ac{EKT} \ac{IE} surface of
\ce{LiH} in Sec.~\ref{SUBSEC:SURFLIH}. In order to speed up the \ac{CASCI}
\ac{EKT} computations in the presence a supplemental diffuse basis function, we
reduced the energy convergence criteria to $10^{-6} \Eh$. The aforementioned
computations were distributed across multiple \acl{HPC} clusters in the \ac{ARC}
center at Virginia Tech.
%=============================================
\section{Results and Discussion}\label{SEC:RESULTS}
%---------------------------------------------
%=============================================

%+++++++++++++++++++++++++++++++++++++++++++++
\subsection{First extended Koopmans's theorem \aclp{IE} of small systems}\label{SUBSEC:ATOMIONIZATION}
%+++++++++++++++++++++++++++++++++++++++++++++
In the following, we investigate the effect of the basis set size, the active
space size, and the diffusivity of the basis functions on the first \ac{EKT}
ionization energies of the \ce{Be} atom, \ce{H-} ion, \ce{H2},
\ce{LiH} and \ce{Li2} molecules.
%~~~~~~~~~~~~~~~~~~~~~~~~~~~~~~~~~~~~~~~~~~~~~
\subsubsection{Beryllium atom}\label{SUBSUBSEC:BE}
%~~~~~~~~~~~~~~~~~~~~~~~~~~~~~~~~~~~~~~~~~~~~~
The spectroscopic properties of \ce{Be} atom have been thoroughly investigated
in the literature both experimentally and
theoretically.\cite{Linstrom:2024:NISTWEBBOOK,Farber:1974:1581,Kelly:1987:1,
  Lide:1992:BOOK,Moore:1970:1,Hildenbrand:1969:5350} We refer to the results of
Refs.~\citenum{Lide:1992:BOOK} and \citenum{Moore:1970:1} for the experimental
value of the \ac{IE} $=9.32263$ eV (or 0.342600 $\Eh$) for the beryllium atom.
Table \ref{TAB:BEIPS} shows the first \ac{EKT} ionization energies of
the \ce{Be} atom calculated using the \ac{CASCI} and \ac{FCI} methods and the
(aug-)cc-pCVXZ basis sets where X = D, T, Q, and
5.\cite{Dunning:1989:1007,Woon:1994:2975,Prascher:2011:69}
\begin{table*}[!htbp]
  \centering
  \setlength{\tabcolsep}{3pt}
  \setlength{\extrarowheight}{1pt}
  \caption{The first \acs{EKT} ground state ionization energy of
    \ce{Be} atom calculated with \acs{CASCI} and \acs{FCI} methods and
    (aug-)cc-pCVXZ basis sets where X = D, T, Q, and 5$^{a}$}
  \label{TAB:BEIPS}
  % \resizebox{\textwidth}{!}{
  \begin{tabular}{lccccc}
    \hline\hline
    Level of Theory                  & X & $\Etot$    & $\IEKTm$ & $\EKTm$  & $\tau_m$$^b$ \\
    \hline
    \mr{4}{*}{CAS(2,4)/cc-pCVXZ}     & D & -14.615452 & 0.348521 & 0.349400 & -0.000879    \\
                                     & T & -14.616531 & 0.348932 & 0.349723 & -0.000791    \\
                                     & Q & -14.616774 & 0.349003 & 0.349788 & -0.000785    \\
                                     & 5 & -14.616832 & 0.349012 & 0.349797 & -0.000785    \\[5pt]
    \mr{4}{*}{CAS(2,4)/aug-cc-pCVXZ} & D & -14.615500 & 0.348490 & 0.349365 & -0.000875    \\
                                     & T & -14.616533 & 0.348927 & 0.349718 & -0.000791    \\
                                     & Q & -14.616775 & 0.349000 & 0.349785 & -0.000785    \\
                                     & 5 & -14.616832 & 0.349010 & 0.349795 & -0.000785    \\[5pt]
    \mr{4}{*}{CAS(4,9)/cc-pCVXZ}     & D & -14.649430 & 0.348522 & 0.348398 & 0.000123     \\
                                     & T & -14.653060 & 0.348943 & 0.348814 & 0.000129     \\
                                     & Q & -14.653807 & 0.349019 & 0.348890 & 0.000129     \\
                                     & 5 & -14.653897 & 0.349028 & 0.348899 & 0.000129     \\[5pt]
    \mr{4}{*}{CAS(4,9)/aug-cc-pCVXZ} & D & -14.649496 & 0.348491 & 0.348366 & 0.000124     \\
                                     & T & -14.653063 & 0.348938 & 0.348809 & 0.000129     \\
                                     & Q & -14.653808 & 0.349016 & 0.348887 & 0.000129     \\
                                     & 5 & -14.653897 & 0.349027 & 0.348898 & 0.000129     \\[5pt]
    \mr{4}{*}{FCI/cc-pCVXZ}          & D & -14.651832 & 0.340953 & 0.340803 & 0.000150     \\
                                     & T & -14.662366 & 0.341922 & 0.341883 & 0.000039     \\
                                     & Q & -14.665680 & 0.342411 & 0.342395 & 0.000016     \\
                                     & 5 & -14.666537 & 0.342507 & 0.342502 & 0.000005     \\
    \hline\hline
  \end{tabular}
  % }
  \begin{tablenotes}
    \item \qquad \qquad \quad $^a$ All energy values are in atomic units.
    \item \qquad \qquad \quad $^b$ The $m$ is either \acs{CASCI} or \acs{FCI}.
  \end{tablenotes}
\end{table*}
Our results highlight the importance of the static correlation effects in
reducing the absolute value of $\TAUCAS$ towards zero as the size of the active
space increases. For instance, adopting the cc-pCV5Z basis set while increasing
the active space size from CAS(2,4) to CAS(4,9) reduces the value of $\TAUCAS$
by approximately 1 $m\Eh$. Note that the CAS(2,4) error values are negative
which violate the inequality \ref{EQ:INEQUALITY} regardless of the adopted basis
set. The reason is attributed\cite{Davidson:2021:051102} to the implementation
details of the \ac{EKT} method in \gamess\ where the domain of the electron
removal operator is not restricted to the active space only. Because in the
conventional implementations of the \ac{EKT}, $k$ in Eq.~\ref{EQ:AK} also
operates on the core orbitals, the resulting hole-space wavefunction becomes
larger than its corresponding \ac{CASCI} counterpart due to the additional
core-related hole \acp{CSF} which should not exist in the \ac{CASCI}
wavefunction.

The results of Table \ref{TAB:BEIPS} also indicate that for a fixed active space
size, augmenting the basis sets with diffuse augmentation functions does not
always reduce the magnitude of $\TAUCAS$. Furthermore, for larger basis sets,
the effect of augmentation with diffuse functions on the error metric,
$\TAUCAS$, diminishes. Although the $\TAUCAS$ approaches a constant value as the
basis set size increases, the \ac{FCI} error metric continuously decreases
towards zero. This is because the diffuse functions of proper symmetry in the
\ac{CASCI} wavefunction are only able to contribute to the target annihilating
orbital through the active space. In contrast, the \ac{FCI} wavefunction can
contain diffuse functions with significant impact on the target annihilating
orbital from the entire orbital space.\cite{Davidson:2021:051102} The best
agreement with the experimental \ac{IE} of the \ce{Be} atom is obtained using
the \ac{FCI} method and the cc-pCV5Z basis set with an absolute error of
0.000093 $\Eh$. Among the \ac{CASCI} \ac{EKT} \acp{IE}, the
CAS(2,4)/aug-cc-pCVDZ level of theory provides the best agreement with the
experiment with an absolute error of 0.005890 $\Eh$.

Table \ref{TAB:BEIPS1S} takes a closer look at the variations of the first
\ac{EKT} ionization energy of the \ce{Be} atom as a supplemental
$s$-type primitive Gaussian function, placed on the nucleus, becomes gradually
diffuse.
\begin{table*}[!htbp]
  \centering
  \setlength{\tabcolsep}{3pt}
  \setlength{\extrarowheight}{1pt}
  \caption{The first \acs{EKT} ground state ionization energy of
    \ce{Be} atom calculated using the \acs{CAS}(4,9) and \acs{FCI} methods and
    cc-pCVDZ basis set augmented by the addition of an atom-centered diffuse
    $s$-type primitive Gaussian function $^{a}$}
  \label{TAB:BEIPS1S}
  % \resizebox{\textwidth}{!}{
  \begin{tabular}{lccccc}
    \hline\hline
    Level of Theory                     & $\al$ & $\Etot$    & $\IEKTm$ & $\EKTm$  & $\tau_m$$^b$ \\
    \hline
    \mr{4}{*}{CAS(4,9)/cc-pCVDZ + $1s$} & 0.010 & -14.649442 & 0.348521 & 0.348397 & 0.000124     \\
                                        & 0.005 & -14.649436 & 0.348538 & 0.348415 & 0.000124     \\
                                        & 0.002 & -14.649432 & 0.348530 & 0.348407 & 0.000123     \\
                                        & 0.001 & -14.649431 & 0.348525 & 0.348402 & 0.000123     \\
    \cline{2-6}
    \mr{1}{*}{CAS(4,9)/cc-pCVDZ}        & --    & -14.649430 & 0.348522 & 0.348398 & 0.000123     \\
    \cline{2-6}
    \mr{4}{*}{FCI/cc-pCVDZ  + $1s$}     & 0.010 & -14.651850 & 0.340775 & 0.340759 & 0.000015     \\
                                        & 0.005 & -14.651841 & 0.340795 & 0.340786 & 0.000009     \\
                                        & 0.002 & -14.651836 & 0.340802 & 0.340799 & 0.000003     \\
                                        & 0.001 & -14.651834 & 0.340805 & 0.340802 & 0.000003     \\
    \cline{2-6}
    \mr{1}{*}{FCI/cc-pCVDZ}             & --    & -14.651832 & 0.340953 & 0.340803 & 0.000150     \\
    \hline\hline
  \end{tabular}
  % }
  \begin{tablenotes}
    \item \qquad \qquad \quad $^a$ All energy values are in atomic units.
    \item \qquad \qquad \quad $^b$ The $m$ is either \acs{CASCI} or \acs{FCI}.
  \end{tablenotes}
\end{table*}
Our results suggest that even using a finite basis set such as cc-pCVDZ, as the
exponent of the primitive function decreases, the $\IEKTFCI$ approaches the
\ac{FCI} ionization energy, $\EKTFCI$, and the error metric, $\TAUFCI$,
approaches zero. Our results corroborate the findings of
Ref.~\citenum{Davidson:2021:051102} where the authors reported similar trends
for the \ce{Be} atom. Due to the restricted nature of the active orbital space
in the \ac{CASCI} wavefunction, the corresponding \ac{EKT} \ac{IE}, $\IEKTCAS$,
appears to be insensitive to the addition of the primitive function and its
diffusivity. In this case, as $\al$ decreases, the error metric $\TAUCAS$
remains constant and positive.
%~~~~~~~~~~~~~~~~~~~~~~~~~~~~~~~~~~~~~~~~~~~~~
\subsubsection{Hydride ion}\label{SUBSUBSEC:HYDRIDE}
%~~~~~~~~~~~~~~~~~~~~~~~~~~~~~~~~~~~~~~~~~~~~~
The hydride ion presents an interesting case study\cite{Rau:1996:113} where the
\ac{EA} of the atomic hydrogen has been measured via tunable-laser threshold
photodetachment spectroscopy to be $6082.99 \pm 0.15\ \wn$ (or approximately
0.0277161 $\Eh$).\cite{Lykke:1991:6104} The usage of the word
``photodetachment'' (instead of photoionization) highlights that the
ground-state wavefunction of the \ce{H-} belongs to a closed-shell anionic
system with a diffuse electron pair while its counterpart, $\Psi^{N-1}$,
corresponds to the neutral hydrogen atom without any Coulomb electron
correlation.

Table \ref{TAB:HYDRIDEIPS} presents the first \ac{EKT} \acp{IE} of the \ce{H-}
ion calculated using the \ac{CASCI} and \ac{FCI} methods and the (aug-)cc-pVXZ
basis sets where X = D, T, Q, and 5.
\begin{table*}[!htbp]
  \centering
  \setlength{\tabcolsep}{3pt}
  \setlength{\extrarowheight}{1pt}
  \caption{The first \acs{EKT} ground state ionization energy of
    \ce{H-} ion calculated with \acs{CASCI} and \acs{FCI} methods and
    (aug-)cc-pVXZ basis sets where X = D, T, Q, and 5$^{a}$}
  \label{TAB:HYDRIDEIPS}
  % \resizebox{\textwidth}{!}{
  \begin{tabular}{lccccc}
    \hline\hline
    Level of Theory                 & X & $\Etot$   & $\IEKTm$  & $\EKTm$   & $\tau_m$$^b$ \\
    \hline
    \mr{4}{*}{CAS(2,2)/cc-pVXZ}     & D & -0.460048 & -0.039231 & -0.039231 & 0.000000     \\
                                    & T & -0.481273 & -0.017452 & -0.017452 & 0.000000     \\
                                    & Q & -0.489744 & -0.008718 & -0.008718 & 0.000000     \\
                                    & 5 & -0.499213 & 0.000751  & 0.000751  & 0.000000     \\[5pt]
    \mr{4}{*}{CAS(2,2)/aug-cc-pVXZ} & D & -0.510886 & 0.012583  & 0.012583  & 0.000000     \\
                                    & T & -0.512342 & 0.013498  & 0.013498  & 0.000000     \\
                                    & Q & -0.512708 & 0.013719  & 0.013719  & 0.000000     \\
                                    & 5 & -0.513094 & 0.014012  & 0.014012  & 0.000000     \\[5pt]
    \mr{4}{*}{CAS(2,5)/cc-pVXZ}     & D & -0.469857 & -0.029422 & -0.029422 & 0.000000     \\
                                    & T & -0.495083 & -0.003705 & -0.003705 & 0.000000     \\
                                    & Q & -0.504529 & 0.005845  & 0.005845  & 0.000000     \\
                                    & 5 & -0.513191 & 0.014496  & 0.014496  & 0.000000     \\[5pt]
    \mr{4}{*}{CAS(2,5)/aug-cc-pVXZ} & D & -0.523061 & 0.024653  & 0.024653  & 0.000000     \\
                                    & T & -0.524550 & 0.025656  & 0.025656  & 0.000000     \\
                                    & Q & -0.524867 & 0.025854  & 0.025854  & 0.000000     \\
                                    & 5 & -0.525073 & 0.025993  & 0.025993  & 0.000000     \\[5pt]
    \mr{4}{*}{FCI/cc-pVXZ}          & D & -0.469857 & -0.029422 & -0.029422 & 0.000000     \\
                                    & T & -0.496312 & -0.003498 & -0.003498 & 0.000000     \\
                                    & Q & -0.506399 & 0.006453  & 0.006453  & 0.000000     \\
                                    & 5 & -0.515313 & 0.015318  & 0.015319  & 0.000000     \\[5pt]
    \mr{4}{*}{FCI/aug-cc-pVXZ}      & D & -0.524029 & 0.024694  & 0.024694  & 0.000000     \\
                                    & T & -0.526562 & 0.026741  & 0.026741  & 0.000000     \\
                                    & Q & -0.527139 & 0.027191  & 0.027191  & 0.000000     \\
                                    & 5 & -0.527429 & 0.027434  & 0.027434  & 0.000000     \\
    \hline\hline
  \end{tabular}
  % }
  \begin{tablenotes}
    \item \qquad \qquad \quad $^a$ All energy values are in atomic units.
    \item \qquad \qquad \quad $^b$ The $m$ is either \acs{CASCI} or \acs{FCI}.
  \end{tablenotes}
\end{table*}
Since the \ac{EKT} is exact for a closed-shell two-electron system, the first
\ac{EKT} \ac{IE} of the \ce{H-} ion, $\IEKTm$ is equal to the energy difference
between the $N$ and $N-1$ states. As such, the error metric, $\tau_m$, where $m$
is either \ac{FCI} or \ac{CASCI}, is exactly zero regardless of the chosen
method, basis set and the active space size.

We should also point out the unphysical negative signs in some of the calculated
\ac{EKT} \acp{IE} (and also the energy differences) for the \ce{H-} ion,
obtained by using the correlation consistent basis sets. This behavior can be
fixed by adopting the augmented versions of the corresponding correlation
consistent basis sets which can sufficiently delineate the diffuse nature of the
electrons in the anionic reference wavefunction. Note that for the cc-pVDZ basis
set, the CAS(2,5) results are equivalent to their \ac{FCI} counterpart within
the same basis set.
%~~~~~~~~~~~~~~~~~~~~~~~~~~~~~~~~~~~~~~~~~~~~~
\subsubsection{Hydrogen molecule}\label{SUBSUBSEC:H2}
%~~~~~~~~~~~~~~~~~~~~~~~~~~~~~~~~~~~~~~~~~~~~~
The hydrogen molecule is a homonuclear diatomic system with a closed-shell
two-electron ground state wavefunction. Similar to the hydride anion case, the
\ac{EKT} is exact for the \ce{H2} molecule. Nonetheless, the reference state is
neutral and the target state is a single-electron cationic state with no
Coulomb electron correlation. There is a rich literature surrounding the
experimental and theoretical studies on the \acp{IE} of \ce{H2} molecule. The
interested reader is referred to the NIST
Webbook\cite{Linstrom:2024:NISTWEBBOOK} for further details. Here, we refer to
the result of McCormack \etal\ who reported the \ac{IE} of the \ce{H2} molecule
to be $15.425932 \pm 0.000002$ eV (or approximately, 0.566892
$\Eh$).\cite{McCormack:1989:2260}

Table \ref{TAB:H2IPS} presents the first \ac{EKT} ionization energies
of the \ce{H2} molecule calculated using the \ac{CASCI} and \ac{FCI} methods and
the (aug-)cc-pVXZ basis sets where X = D, T, Q, and 5.
\begin{table*}[!htbp]
  \centering
  \setlength{\tabcolsep}{3pt}
  \setlength{\extrarowheight}{1pt}
  \caption{The first \acs{EKT} ground state ionization energy of
    \ce{H2} molecule calculated with \acs{CASCI} and \acs{FCI} methods and
    (aug-)cc-pVXZ basis sets where X = D, T, Q, and 5$^{a}$}
  \label{TAB:H2IPS}
  % \resizebox{\textwidth}{!}{
  \begin{tabular}{lccccc}
    \hline\hline
    Level of Theory                  & X & $\Etot$   & $\IEKTm$ & $\EKTm$  & $\tau_m$$^b$ \\
    \hline
    \mr{4}{*}{CAS(2,2)/cc-pVXZ}      & D & -1.146874 & 0.608888 & 0.608888 & 0.000000     \\
                                     & T & -1.151403 & 0.611338 & 0.611338 & 0.000000     \\
                                     & Q & -1.151962 & 0.611558 & 0.611558 & 0.000000     \\
                                     & 5 & -1.152123 & 0.611592 & 0.611592 & 0.000000     \\[5pt]
    \mr{4}{*}{CAS(2,2)/aug-cc-pVXZ}  & D & -1.146951 & 0.609301 & 0.609301 & 0.000000     \\
                                     & T & -1.151457 & 0.611300 & 0.611300 & 0.000000     \\
                                     & Q & -1.151974 & 0.611546 & 0.611546 & 0.000000     \\
                                     & 5 & -1.152125 & 0.611592 & 0.611592 & 0.000000     \\[5pt]
    \mr{4}{*}{CAS(2,10)/cc-pVXZ}     & D & -1.163374 & 0.598173 & 0.598173 & 0.000000     \\
                                     & T & -1.170821 & 0.604730 & 0.604730 & 0.000000     \\
                                     & Q & -1.171882 & 0.603115 & 0.603115 & 0.000000     \\
                                     & 5 & -1.172092 & 0.603115 & 0.603115 & 0.000000     \\[5pt]
    \mr{4}{*}{CAS(2,10)/aug-cc-pVXZ} & D & -1.164411 & 0.599291 & 0.599291 & 0.000000     \\
                                     & T & -1.171191 & 0.602471 & 0.602471 & 0.000000     \\
                                     & Q & -1.171908 & 0.603099 & 0.603099 & 0.000000     \\
                                     & 5 & -1.172096 & 0.603121 & 0.603121 & 0.000000     \\[5pt]
    \mr{4}{*}{FCI/cc-pVXZ}           & D & -1.163374 & 0.598173 & 0.598173 & 0.000000     \\
                                     & T & -1.172332 & 0.603191 & 0.603191 & 0.000000     \\
                                     & Q & -1.173794 & 0.604198 & 0.604198 & 0.000000     \\
                                     & 5 & -1.174221 & 0.604490 & 0.604489 & 0.000001     \\
    \hline\hline
  \end{tabular}
  % }
  \begin{tablenotes}
    \item \qquad \qquad \quad $^a$ All energy values are in atomic units.
    \item \qquad \qquad \quad $^b$ The $m$ is either \acs{CASCI} or \acs{FCI}.
  \end{tablenotes}
\end{table*}
Similar to the results of Table \ref{TAB:HYDRIDEIPS}, the first \ac{EKT}
ionization energies of the \ce{H2} molecule are exact for the
\ac{CASCI} and \ac{FCI} methods regardless of the adopted basis set and the
active space sizes. Therefore, the \ac{CASCI} and \ac{FCI} error metrics are
exactly zero for all cases except that of FCI/cc-pV5Z which is affected by the
rounding error. The best agreement with the experimental \ac{IE} of the \ce{H2}
molecule is simultaneously obtained by the \ac{CASCI}/cc-pVDZ and
\ac{FCI}/cc-pVDZ levels of theory with an absolute error value of 0.031281
$\Eh$.
%~~~~~~~~~~~~~~~~~~~~~~~~~~~~~~~~~~~~~~~~~~~~~
\subsubsection{Lithium hydride}\label{SUBSUBSEC:LIH}
%~~~~~~~~~~~~~~~~~~~~~~~~~~~~~~~~~~~~~~~~~~~~~
Lithium hydride is a heteronuclear diatomic molecule with a closed-shell
four-electron ground state wavefunction of $^1\Sigma$ symmetry. The experimental
value of the \ac{IE} for the \ce{LiH} molecule is $7.9 \pm 0.3$ eV (or
approximately 0.2903 $\Eh$).\cite{Linstrom:2024:NISTWEBBOOK}

Table \ref{TAB:LIHIPS} presents the first \acs{EKT} ground state acp{IE} of the
\ce{LiH} molecule calculated with \acs{CASCI} and \acs{FCI} methods and
(aug-)cc-pCVXZ basis sets where X = D, T, Q, and 5.
\begin{table*}[!htbp]
  \centering
  \setlength{\tabcolsep}{3pt}
  \setlength{\extrarowheight}{1pt}
  \caption{The first \acs{EKT} ground state ionization energy of
    \ce{LiH} molecule calculated with \acs{CASCI} and \acs{FCI} methods and
    (aug-)cc-pCVXZ basis sets where X = D, T, Q, and 5$^{a}$}
  \label{TAB:LIHIPS}
  % \resizebox{\textwidth}{!}{
  \begin{tabular}{lccccc}
    \hline\hline
    Level of Theory                  & X & $\Etot$   & $\IEKTm$ & $\EKTm$  & $\tau_m$$^b$ \\
    \hline
    \mr{4}{*}{CAS(2,2)/cc-pCVXZ}     & D & -8.000497 & 0.278143 & 0.278151 & -0.000008    \\
                                     & T & -8.003121 & 0.278249 & 0.278259 & -0.000010    \\
                                     & Q & -8.003593 & 0.278714 & 0.278726 & -0.000012    \\
                                     & 5 & -8.003711 & 0.278718 & 0.278731 & -0.000012    \\[5pt]
    \mr{4}{*}{CAS(2,2)/aug-cc-pCVXZ} & D & -8.000921 & 0.278600 & 0.278608 & -0.000007    \\
                                     & T & -8.003236 & 0.278425 & 0.278436 & -0.000011    \\
                                     & Q & -8.003614 & 0.278714 & 0.278726 & -0.000012    \\
                                     & 5 & -8.003715 & 0.278718 & 0.278730 & -0.000012    \\[5pt]
    \mr{4}{*}{CAS(2,5)/cc-pCVXZ}     & D & -8.013851 & 0.290370 & 0.290587 & -0.000217    \\
                                     & T & -8.020334 & 0.293441 & 0.293729 & -0.000287    \\
                                     & Q & -8.021076 & 0.294016 & 0.294303 & -0.000287    \\
                                     & 5 & -8.021234 & 0.294067 & 0.294356 & -0.000290    \\[5pt]
    \mr{4}{*}{CAS(2,5)/aug-cc-pCVXZ} & D & -8.016722 & 0.292778 & 0.292930 & -0.000152    \\
                                     & T & -8.020658 & 0.293718 & 0.293995 & -0.000277    \\
                                     & Q & -8.021130 & 0.294048 & 0.294335 & -0.000287    \\
                                     & 5 & -8.021246 & 0.294072 & 0.294362 & -0.000289    \\[5pt]
    \mr{4}{*}{FCI/cc-pCVXZ}          & D & -8.048672 & 0.289338 & 0.289313 & 0.000025     \\
                                     & T & -8.064711 & 0.293986 & 0.293980 & 0.000006     \\
                                     & Q & --        & --       & --       & --           \\
                                     & 5 & --        & --       & --       & --           \\
    \hline\hline
  \end{tabular}
  % }
  \begin{tablenotes}
    \item \qquad \qquad \quad $^a$ All energy values are in atomic units.
    \item \qquad \qquad \quad $^b$ The $m$ is either \acs{CASCI} or \acs{FCI}.
  \end{tablenotes}
\end{table*}
The results of Table \ref{TAB:LIHIPS} demonstrate that for the CAS(2,2)
reference wavefunction, increasing the size of the basis set from cc-pCVDZ to
cc-pCV5Z does not improve the accuracy of the first \ac{CASCI} \ac{EKT}
ionization energies and in fact, slightly increases the magnitude of the error
metric $\TAUCAS$. The addition of augmentation functions to the basis set, in
going from the cc-pCVXZ to the aug-cc-pCVXZ basis sets, does not significantly
change the error metric $\TAUCAS$ for the CAS(2,2) wavefunction. Adding three
$2p$ orbitals of the lithium atom to the active space in the CAS(2,5)
wavefunction will increase the absolute error by more than an order of magnitude
from $-0.000012$ to $-0.000290$ $\Eh$ within the cc-pCV5Z basis and to
$-0.000289$ $\Eh$ in the aug-cc-pCV5Z. The addition of augmentation diffuse
functions to the basis set going from cc-pCVXZ to aug-cc-pCVXZ basis sets
improves the accuracy of the first \ac{EKT} ionization energies for the CAS(2,5)
wavefunction expressed within the  double- and triple-zeta basis sets but leaves
the error values pertinent to the quadruple- and quintuple-zeta basis sets
almost unchanged.

Similar to the case of CAS(2,4) wavefunction for \ce{Be} atom in Table
\ref{TAB:BEIPS}, the negative signs of the error metric, $\TAUCAS$, in Table
\ref{TAB:LIHIPS} are in violation of the inequality \ref{EQ:INEQUALITY} as the
domain of the electron removal operator in Eq.~\ref{EQ:AK} is not restricted to
the active space only. This problem is automatically resolved for the \ac{FCI}
wavefunction where the entire orbital space contributes to the ionization
process. Table \ref{TAB:LIHIPS} shows that the error metric $\TAUFCI$ is
positive and its magnitude decreases towards zero as the size of the basis set
increases. Unfortunately, the \ac{FCI} calculations for the cc-pCVQZ and
cc-pCV5Z basis sets could not be carried out due to the computational cost
barrier.
%~~~~~~~~~~~~~~~~~~~~~~~~~~~~~~~~~~~~~~~~~~~~~
\subsubsection{Lithium dimer}\label{SUBSUBSEC:LI2}
%~~~~~~~~~~~~~~~~~~~~~~~~~~~~~~~~~~~~~~~~~~~~~
Lithium dimer is a six-electron closed-shell homonuclear diatomic molecule with
a ground state wavefunction of $^1\Sigma_g$ symmetry. The \ac{FCI} \ac{EKT}
wavefunction is not exact except at the complete basis set limit. The reference
state is neutral and its ($N\!-\!1$)-electron counterpart is cationic. The
experimental \ac{IE} value of 5.1127 $\pm$ 0.0003 eV (or 0.187888 $\Eh$) is
taken from an optical spectroscopy measurement in the
literature.\cite{McGeoch:1983:347,Linstrom:2024:NISTWEBBOOK}

Table \ref{TAB:LI2IPS} presents the first \ac{EKT} ionization energies
of the \ce{Li2} molecule calculated using the \ac{CASCI} and \ac{FCI} methods
and the (aug-)cc-pCVXZ basis sets where X = D, T, Q, and 5.
\begin{table*}[!htbp]
  \centering
  \setlength{\tabcolsep}{3pt}
  \setlength{\extrarowheight}{1pt}
  \caption{The first \acs{EKT} ground state ionization energy of
    \ce{Li2} molecule calculated with the \acs{CASCI} method and
    (aug-)cc-pCVXZ basis sets where X = D, T, Q, and 5$^{a}$}
  \label{TAB:LI2IPS}
  % \resizebox{\textwidth}{!}{
  \begin{tabular}{lccccc}
    \hline\hline
    Level of Theory                  & X & $\Etot$    & $\IEKTCAS$ & $\EKTCASCI$ & $\TAUCAS$ \\
    \hline
    \mr{4}{*}{CAS(2,2)/cc-pCVXZ}     & D & -14.879357 & 0.192402   & 0.192445    & -0.000043 \\
                                     & T & -14.880318 & 0.192456   & 0.192506    & -0.000050 \\
                                     & Q & -14.884197 & 0.192933   & 0.193031    & -0.000097 \\
                                     & 5 & -14.884264 & 0.192951   & 0.193048    & -0.000097 \\[5pt]
    \mr{4}{*}{CAS(2,2)/aug-cc-pCVXZ} & D & -14.879378 & 0.192445   & 0.192489    & -0.000043 \\
                                     & T & -14.880347 & 0.192544   & 0.192594    & -0.000051 \\
                                     & Q & -14.884208 & 0.192955   & 0.193052    & -0.000097 \\
                                     & 5 & -14.884268 & 0.192958   & 0.193055    & -0.000097 \\[5pt]
    \mr{4}{*}{CAS(2,8)/cc-pCVXZ}     & D & -14.901514 & 0.191527   & 0.191660    & -0.000133 \\
                                     & T & -14.903058 & 0.191711   & 0.191842    & -0.000132 \\
                                     & Q & -14.903319 & 0.191839   & 0.191966    & -0.000127 \\
                                     & 5 & -14.903384 & 0.191844   & 0.191971    & -0.000127 \\[5pt]
    \mr{4}{*}{CAS(2,8)/aug-cc-pCVXZ} & D & -14.901580 & 0.191568   & 0.191699    & -0.000131 \\
                                     & T & -14.903116 & 0.191722   & 0.191852    & -0.000130 \\
                                     & Q & -14.903324 & 0.191842   & 0.191969    & -0.000127 \\
                                     & 5 & -14.903386 & 0.191845   & 0.191972    & -0.000127 \\
    \hline\hline
  \end{tabular}
  % }
  \begin{tablenotes}
    \item \qquad \qquad \quad $^a$ All energy values are in atomic units.
  \end{tablenotes}
\end{table*}
According to the results of Table \ref{TAB:LI2IPS}, the CAS(2,2) wavefunction,
involving the $2\sig$ and $2\sig^*$ molecular orbitals of the lithium dimer in
the active space, yields more accurate \ac{EKT} \acp{IE} than its CAS(2,8)
counterpart for all adopted basis sets. In particular, by including all six
additional $3\sig$, $3\sig^*$,$1\pi$ and $1\pi^*$ molecular orbitals in the
active space, the magnitude of the absolute error $\TAUCAS$ almost doubles using
the triple-zeta basis. Our results also suggest that adding the augmentation
functions to the basis set does not significantly improve the accuracy
of the \ac{EKT} \acp{IE} for the CAS(2,2) and CAS(2,8) wavefunctions.

Similar to the \ce{LiH} results in Table \ref{TAB:LIHIPS}, the calculated values
of the \ac{CASCI} error metric, $\TAUCAS$, for \ce{Li2} in Table \ref{TAB:LI2IPS}
are also negative for all studied active space and basis set size combinations.
The calculation of \ac{FCI} \ac{EKT} \acp{IE} for the \ce{Li2} molecule was not
computationally feasible due to the large number of determinants in the
wavefunction expansion.
%+++++++++++++++++++++++++++++++++++++++++++++
\subsection{Analysis of Ernzerhof's Conjecture for \ce{LiH}}\label{SUBSEC:SURFLIH}
%+++++++++++++++++++++++++++++++++++++++++++++
Ernzerhof's conjecture\cite{Ernzerhof:2009:793} has crucial implications on
the choice of the basis set and the numerical accuracy of the \ac{EKT} \acp{IE}.
The conjecture states that the exact lowest \ac{IE} of an atom or molecule can
be approximated by the \ac{EKT} model with arbitrary accuracy if the adopted
basis set includes a diffuse basis function of appropriate symmetry which allows
an electron to travel very far from the system. As such, the approximate
hole-state wavefunction for the ($N\!-\!1$)-electron entity can asymptotically
approach its exact counterpart as the distance of the electron approaches
infinity.

\begin{figure}[!tbph]
  \centering
  \setlength{\abovecaptionskip}{-2pt}
  \includegraphics[]{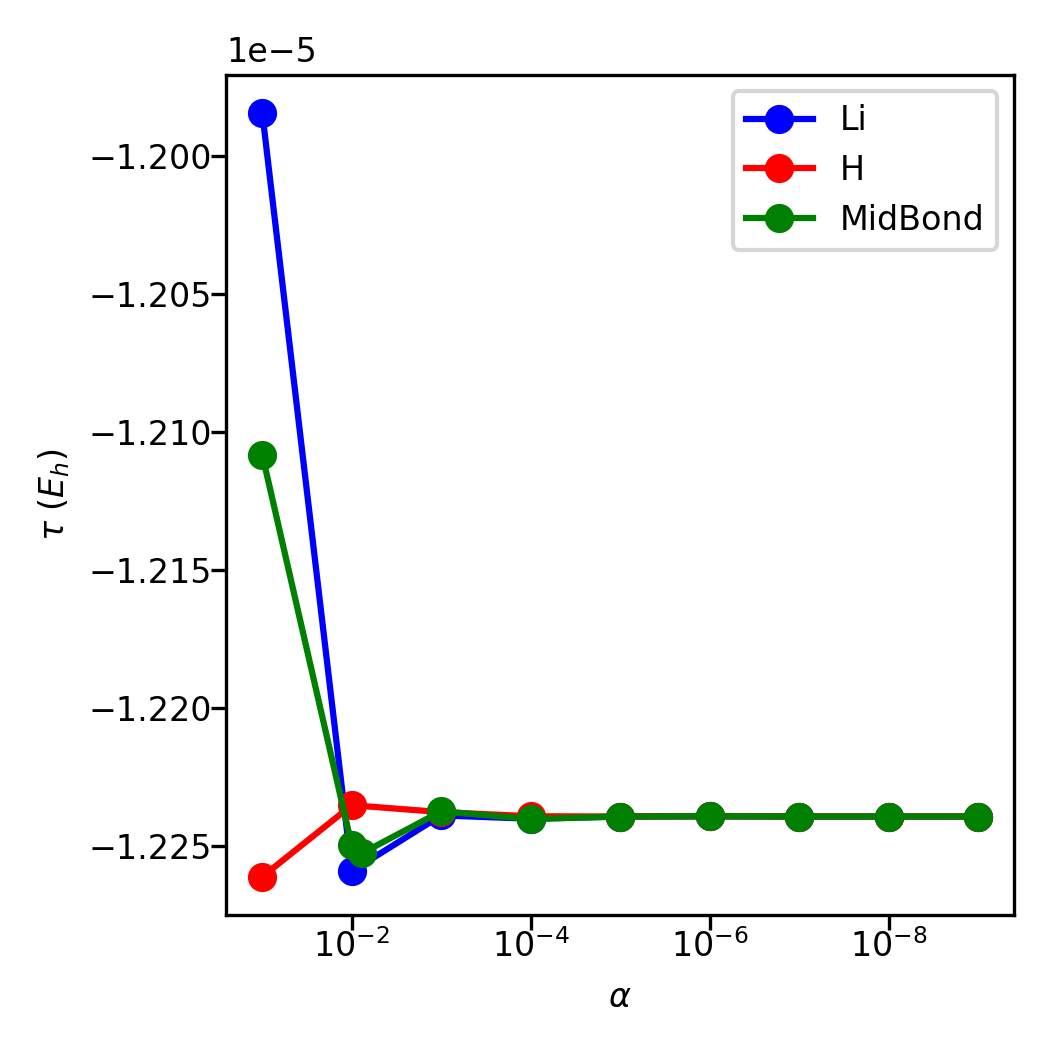}
  \caption{Variations of $\TAUCAS$ for \ce{LiH} as a function of the exponent
    $\al$ and the mean position $\br_0$ of a $s$-type primitive Gaussian
    function of the form $\phi_s(\br) \propto \exp[-\al (\br-\br_0)^2]$ when
    placed on the \ce{Li} (blue), \ce{H} (red) or the mid-bond (green).
    Here, $R = \Req=1.5949$ \AA.}
  \label{FIG:ALPHASCAN}
\end{figure}

Figure \ref{FIG:ALPHASCAN} demonstrates the results of our investigation on the
variations of the numerical error in the \ac{CASCI} \ac{EKT} ionization
energies, $\TAUCAS$, of \ce{LiH} molecule as a function of the exponent, $\al$,
and the reference position, $\br_0$, of a supplemental $s$-type primitive
Gaussian function of the form $\phi_s(\br) \propto \exp[-\al (\br-\br_0)^2]$.
The \ac{EKT} calculations are performed at the CAS(2,2)/aug-cc-pwCVTZ level of
theory and the internuclear distance of $R = \Req = 1.5949$ \AA. We place the
primitive Gaussian function on the \ce{Li} atom, \ce{H} atom or the mid-bond,
and each time, change the exponent values between $10^{-1}$ and $10^{-9}$ with a
step size of $10^{-1}$. As Fig.~\ref{FIG:ALPHASCAN} shows, for large values of
$\al$ ($>10^{-3}$), the reference position of the augmentation Gaussian function
has a small impact ($< 1 \mu \Eh$) on the \ac{CASCI} \ac{EKT} ionization energy
and the error metric, $\TAUCAS$. Nonetheless, as $\al$ decreases, the impact of
the reference position of the augmentation function on the \ac{CASCI} \ac{EKT}
ionization energies diminishes. As a result, both calculated \ac{CASCI} \ac{EKT}
ionization energy and the error metric $\TAUCAS$ converge to a constant value as
$\al \rightarrow 0$.

\begin{figure}[!tbph]
  \centering
  \setlength{\abovecaptionskip}{-3pt}
  \includegraphics{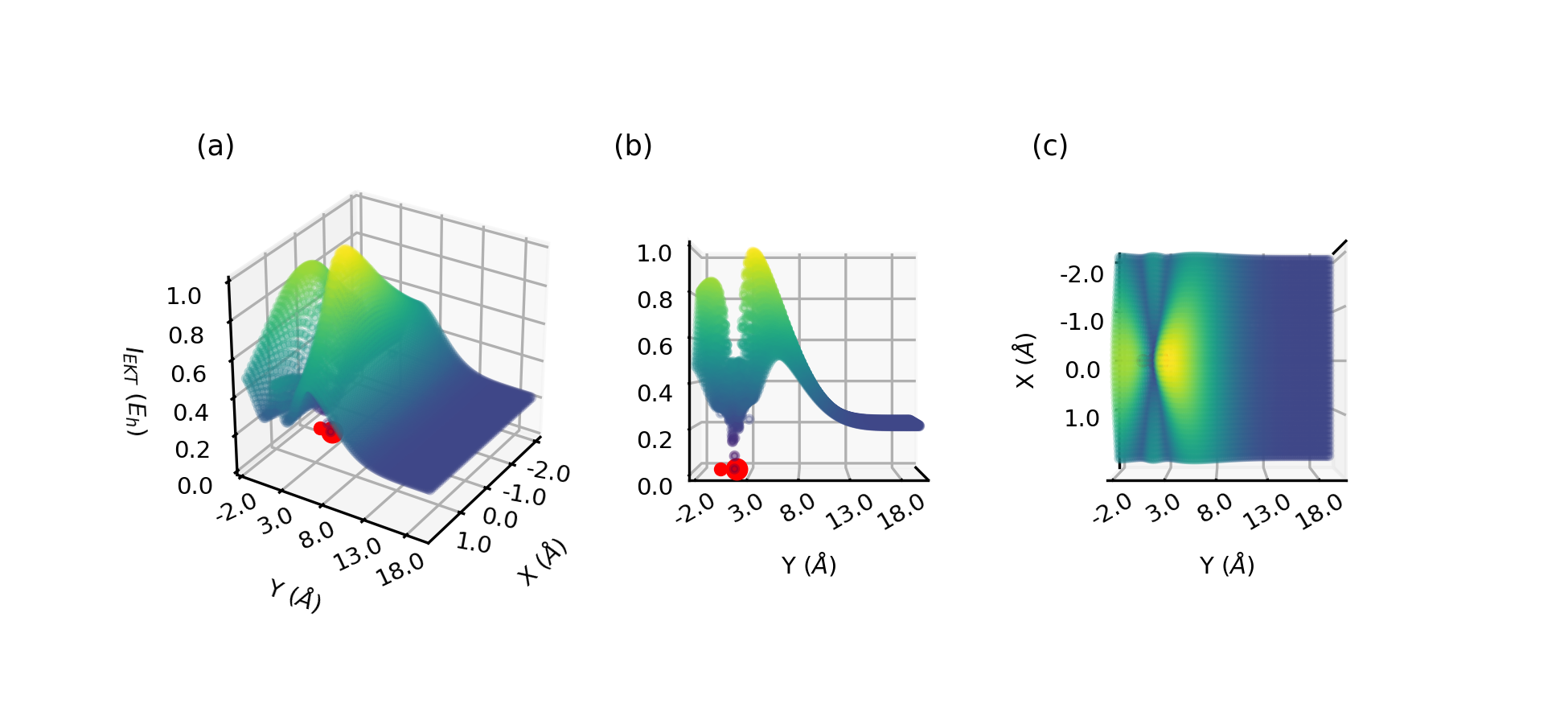}
  \caption{The first \ac{EKT} ionization energy surface of \ce{LiH}
    molecule calculated at the CAS(2,2)/aug-cc-pwCVTZ level of theory. The
    bond length is fixed at $R=\Req=1.5949$ \AA.}
  \label{FIG:IEKT3D}
\end{figure}

We have also investigated the role of supplemental Gaussian function's reference
point in Ernzerhof's conjecture by creating a three-dimensional \ac{CASCI}
\ac{EKT} ionization energy surface for \ce{LiH} molecule at the
CAS(2,2)/aug-cc-pwCVTZ level of theory (Fig.~\ref{FIG:IEKT3D}).
The generation of the \ac{IE} surface is a computationally demanding multi-step
process and its details can be found in the \si. The energy values in
Fig.~\ref{FIG:IEKT3D} are normalized to allow for a consistent representation
and comparison within a unified coordinate system across all studied
internuclear distances of $R=\Req$, $5\Req$, and $15\Req$ where $\Req=1.5949$
\AA. Figure \ref{FIG:IEKT3D} demonstrates that the \ac{EKT} ionization energy
values become smaller around the middle bond and closer to the \ce{Li} atom
compared with those in other regions of space. The \ac{EKT} \ac{IE} surface also
exhibits multiple minima and two individual maxima around the \ce{H} and \ce{Li}
atoms the latter being more pronounced. As one moves away from the molecule
along the $z$-axis, the \ac{EKT} ionization energy values rapidly decay to a
constant value which is consistent with our previous observations for the
$\TAUCAS$ values in Tables~\ref{TAB:BEIPS}--\ref{TAB:LIHIPS}.
%=============================================
\section{Conclusion}\label{SEC:CONCLUSION}
%---------------------------------------------
In this manuscript, we have investigated the impacts of the active space and
basis set size on the accuracy of the \ac{CASCI} and \ac{FCI} \ac{EKT}
ionization energies of a variety of atomic and molecular systems
within the context of Ernzerhof's conjecture.\cite{Ernzerhof:2009:793} In
discussing our results, it is important to distinguish between the finite and
complete basis set regimes. In the complete basis set limit, the \ac{HCI} and
\ac{FCI} wavefunctions become equivalent\cite{Davidson:2021:051102} as the
energy of ($N\!-\!1$) electron state wavefunction approaches its exact value.
Our results also demonstrate that in a truncated (or complete) basis set, the
\ac{CASCI} wavefunction cannot become equivalent to its \ac{HCI} counterpart if
the domain of $k$ in the electron removal operator $\hat{A}_k$
(Eq.~\ref{EQ:AK}) is not restricted to the active orbital
space.\cite{Davidson:2021:051102} This issue plagues several modern quantum
chemistry codes, including \gamess\, because of which the inequality in
Eq.~\ref{EQ:INEQUALITY} can be violated. Furthermore, for systems with more than
two electrons, no electron removal from the reference $N$-electron wavefunction
can yield the corresponding exact ($N\!-\!1$)-electron wavefunction. As such, if
$N>2$, the first \ac{FCI} \ac{EKT} ionization energy cannot be equal to the
energy difference between the $N$-electron reference wavefunction and its
($N\!-\!1$)-electron counterpart. Nevertheless, in a finite basis set, according
to Ernzerhof's conjecture,\cite{Ernzerhof:2009:793} $\IEKTFCI$ can get
arbitrarily close to $\EKTFCI$ if the basis set contains a function that can
have a major contribution to the reference $N$-electron wavefunction but does
not significantly impact the ($N\!-\!1$)-electron
wavefunction.\cite{Davidson:2021:051102,Ernzerhof:2009:793}

Our numerical results for the \ce{Be} atom, \ce{H-}, \ce{H2}, \ce{LiH}, and
\ce{Li2} molecules support Ernzerhof's conjecture for the \ac{FCI} reference
wavefunction. However, the analysis of the \ac{CASCI} wavefunction represented
in an incomplete basis set is more complex. Our investigations on the \ac{EKT}
\acp{IE} of the \ce{LiH} molecule suggest that the first \ac{CASCI} \ac{EKT}
ionization energy cannot be exact due to the limited impact of supplementary
diffuse primitive Gaussian function to the target annihilated orbital only
through the active orbital set. The reference position of the supplementary
Gaussian function can have a small effect on the \ac{CASCI} \ac{EKT} ionization
energy for larger values of the exponent, $\al$. However, as $\al$ decreases,
the impact of the reference position of the supplementary Gaussian function on
the \ac{CASCI} \ac{EKT} ionization energies diminishes. Similar trends have also
been observed for the \ac{EKT} \ac{IE} surface of the \ce{LiH} molecule. Our
results suggest that the \ac{CASCI} \ac{EKT} ionization energy values are
smaller around the middle bond and closer to the \ce{Li} atom than those in
other regions of space. The \ac{EKT} \ac{IE} surface also exhibits multiple
minima and two individual maxima around the \ce{H} and \ce{Li} atoms the latter
being more pronounced. As the supplemental Gaussian function moves away from the
molecule along the bond axis, the \ac{EKT} ionization energy values rapidly
decay to a constant value which is consistent with our previous observations for
the $\TAUCAS$ values in Tables~\ref{TAB:BEIPS}--\ref{TAB:LIHIPS}.
%=============================================
%
%=================================
%
%
%
%=================================
% Acknowledgements
%---------------------------------
\acresetall	% reset all acronyms
%---------------------------------
\section{Acknowledgements}
This work is funded by the \acl{NSF} grant CHE-2136142 and supported by the Advanced Research 
Computing (https://arc.vt.edu) at Virginia Tech which provided computational resources and 
technical assistance that have contributed to the results of this manuscript.
RH would like to thank Prof. Viktor Staroverov for helpful discussions. MM is grateful to 
NVIDIA Corporation for the generous Academic Hardware Grant.
%=================================

\newpage

%\section{Acronyms}
\begin{acronym}
    \acrodefplural{1-RDM}{one-electron \aclp{RDM}}
    \acrodefplural{2-RDM}{two-electron \aclp{RDM}}
    \acrodefplural{3-RDM}{three-electron \aclp{RDM}}
    \acrodefplural{4-RDM}{four-electron \aclp{RDM}}
    \acrodefplural{EA}{electron affinities}
    \acrodefplural{IE}{ionization energies}
    \acrodefplural{RDM}{reduced density matrices}
    \acro{1-HRDM}{one-hole \acl{RDM}}
    \acro{1-RDM}{one-electron \acl{RDM}}
    \acrodef{1H-PDFT}{one-parameter hybrid \acl{PDFT}}
    \acro{2-HRDM}{two-hole \acl{RDM}}
    \acro{2-RDM}{two-electron \acl{RDM}}
    \acro{3-RDM}{three-electron \acl{RDM}}
    \acro{4-RDM}{four-electron \acl{RDM}}
    \acro{ACI}{adaptive \acl{CI}}
    \acro{ACI-DSRG-MRPT2}{\acl{ACI}-\acl{DSRG} \acl{MR} \acl{PT2}}
    \acro{ACSE}{anti-Hermitian \acl{CSE}}
    \acrodef{AKEEPE}[$\AKEEe$]{absolute kinetic energy error per electron}
    \acrodef{AKEE}{absolute kinetic energy error}
    \acro{AO}{atomic orbital}
    \acro{AQCC}{averaged quadratic \acl{CC}}
    \acro{ARC}{Advanced Research Computing}
    \acro{aug-cc-pVQZ}{augmented correlation-consistent polarized-valence quadruple-$\ze$}
    \acro{aug-cc-pVTZ}{augmented correlation-consistent polarized-valence triple-$\ze$}
    \acro{aug-cc-pwCV5Z}[aug-cc-pwCV5Z]{augmented correlation-consistent polarized weighted core-valence quintuple-$\ze$}
    \acro{B3LYP}{Becke-3-\acl{LYP}}
    \acro{BLA}{bond length alternation}
    \acro{BLYP}{Becke and \acl{LYP}}
    \acro{BO}{Born-Oppenheimer}
    \acro{BP86}{Becke 88 exchange and P86 Perdew-Wang correlation}
    \acro{BPSDP}{boundary-point \acl{SDP}}
    \acro{CAM}{Coulomb-attenuating method}
    \acro{CAM-B3LYP}{Coulomb-attenuating method \acl{B3LYP}}
    \acro{CASCI}[CAS-CI]{\acl{CAS} \acl{CI}}
    \acro{CAS-PDFT}{\acl{CAS} \acl{PDFT}}
    \acro{CASPT2}{\acl{CAS} \acl{PT2}}
    \acro{CASSCF}{\acl{CAS} \acl{SCF}}
    \acro{CAS}{complete active-space}
    \acro{cc-pVDZ}{correlation-consistent polarized-valence double-$\ze$}
    \acro{cc-pVTZ}{correlation-consistent polarized-valence triple-$\ze$}
    \acro{CCSDT}{coupled-cluster, singles doubles and triples}
    \acro{CCSD}{coupled-cluster with singles and doubles}
    \acro{CC}{coupled-cluster}
    \acro{CI}{configuration interaction}
    \acro{CO}{constant-order}
    \acro{CPO}{correlated participating orbitals}
    \acro{CSF}{configuration state function}
    \acro{CS-KSDFT}{\acl{CS}-\acl{KSDFT}}
    \acro{CSE}{contracted \acl{SE}}
    \acro{CS}{constrained search}
    \acro{DC-DFT}{density corrected-\acl{DFT}}
    \acro{DC-KDFT}{density corrected-\acl{KDFT}}
    \acro{DE}{delocalization error}
    \acro{DFT}{density functional theory}
    \acro{DF}{density-fitting}
    \acro{DIIS}{direct inversion in the iterative subspace}
    \acro{DMRG}{density matrix renormalization group}
    \acro{DOCI}{doubly occupied \acl{CI}}
    \acro{DSRG}{driven similarity renormalization group}
    \acro{EA}{electron affinity}
    \acro{EKT}{extended Koopmans's theorem}
    \acro{ERI}{electron-repulsion integral}
    \acro{EUE}{effectively unpaired electron}
    \acro{FC}{fractional calculus}
    \acro{FCI}{full \acl{CI}}
    \acro{FP-1}{frontier partition with one set of interspace excitations}
    \acro{FSE}{fractional \acl{SE}}
    \acro{ftBLYP}{fully \acl{tBLYP}}
    \acro{ftPBE}{fully \acl{tPBE}}
    \acro{ftSVWN3}{fully \acl{tSVWN3}}
    \acro{ft}{full translation}
    \acro{GASSCF}{generalized active-space \acl{SCF}}
    \acro{GGA}{generalized gradient approximation}
    \acro{GL}{Gr\"unwald-Letnikov}
    \acro{GMCPDFT}[G-MC-PDFT]{generalized \acl{MCPDFT}}
    \acro{GTO}{Gaussian-type orbital}
    \acro{HCI}{hole-state \acl{CI}}
    \acro{HF}{Hartree-Fock}
    \acro{HISS}{Henderson-Izmaylov-Scuseria-Savin}
    \acrodef{HK}{Hohenberg-Kohn}
    \acro{HOMO}{highest-occupied \acl{MO}}
    \acro{HONO}{highest-occupied \acl{NO}}
    \acro{HPC}{high-performance computing}
    \acro{HPDFT}{hybrid \acl{PDFT}}
    \acro{HRDM}{hole \acl{RDM}}
    \acro{HSE}{Heyd-Scuseria-Ernzerhof}
    \acro{HXC}{Hartree-\acl{XC}}
    \acro{IE}{ionization energy}
    \acro{IPEA}{ionization potential electron affinity}
    \acro{IPSDP}{interior-point \acl{SDP}}
    \acro{KDFT}{kinetic \acl{DFT}}
    \acro{KSDFT}[KS-DFT]{\acl{KS} \acl{DFT}}
    \acro{KS}{Kohn-Sham}
    \acro{KT}{Koopmans's theorem}
    \acro{LBFGS}[L-BFGS]{limited-memory Broyden-Fletcher-Goldfarb-Shanno}
    \acro{LC}{long-range corrected}
    \acro{LC-VV10}{\acl{LC} Vydrov-van Voorhis 10}
    \acro{l-DFVB}[$\la$-DFVB]{$\la$-density functional \acl{VB}}
    \acro{LEB}{local energy balance}
    \acro{LE}{localization error}
    \acrodef{lftBLYP}[$\la$-\acs{ftBLYP}]{$\la$-\acl{ftBLYP}}
    \acrodef{lftPBE}[$\la$-\acs{ftPBE}]{$\la$-\acl{ftPBE}}
    \acrodef{lftrevPBE}[$\la$-\acs{ftrevPBE}]{$\la$-\acl{ftrevPBE}}
    \acrodef{lftSVWN3}[$\la$-\acs{ftSVWN3}]{$\la$-\acl{ftSVWN3}}
    \acro{lMCPDFT}[$\la$-MC-PDFT]{\acl{MC} \acl{1H-PDFT}}
    \acro{LMF}{local mixing function}
    \acro{LO}{Lieb-Oxford}
    \acro{LP}{linear programming}
    \acro{LR}{long-range}
    \acro{LSDA}{local spin-density approximation}
    \acrodef{ltBLYP}[$\la$-\acs{tBLYP}]{$\la$-\acl{tBLYP}}
    \acrodef{ltPBE}[$\la$-\acs{tPBE}]{$\la$-\acl{tPBE}}
    \acrodef{ltrevPBE}[$\la$-\acs{trevPBE}]{$\la$-\acl{trevPBE}}
    \acrodef{ltSVWN3}[$\la$-\acs{tSVWN3}]{$\la$-\acl{tSVWN3}}
    \acro{LUMO}{lowest-unoccupied \acl{MO}}
    \acro{LUNO}{lowest-unoccupied \acl{NO}}
    \acro{LYP}{Lee-Yang-Parr}
    \acro{M06}{Minnesota 06}
    \acro{M06-2X}{\acl{M06} with double non-local exchange}
    \acro{M06-L}{\acl{M06} local}
    \acro{MAEPE}[$\MAEe$]{\acl{MAE} per electron}
    \acro{MAE}{mean absolute error}
    \acro{MAKEE}[$\MAKEE$]{mean absolute kinetic energy error per electron}
    \acro{MAX}{maximum absolute error}
    \acro{MC1H-PDFT}{\acl{MC} \acl{1H-PDFT}}
    \acro{MC1H}{\acl{MC} one-parameter hybrid \acl{PDFT}}
    \acro{MCHPDFT}{\acl{MC} hybrid-\acl{PDFT}}
    \acro{MCPDFT}[MC-PDFT]{\acl{MC} \acl{PDFT}}
    \acro{MCRSHPDFT}[$\mu\la$-MCPDFT]{\acl{MC} range-separated hybrid-\acl{PDFT}}
    \acro{MCSCF}{\acl{MC} \acl{SCF}}
    \acrodef{MC}{multiconfiguration}
    \acro{MN15}{Minnesota 15}
    \acro{MOLSSI}[MolSSI]{Molecular Sciences Software Institute}
    \acro{MO}{molecular orbital}
    \acro{MP2}{second-order M\o ller-Plesset \acl{PT}}
    \acro{MR-AQCC}{\acl{MR}-averaged quadratic \acl{CC}}
    \acrodef{MR}{multireference}
    \acro{MS0}{MS0 meta-GGA exchange and revTPSS GGA correlation}
    \acro{NDDO}{neglect of diatomic differential overlap}
    \acro{NGA}{nonseparable gradient approximation}
    \acro{NIAD}{normed integral absolute deviation}
    \acro{NO}{natural orbital}
    \acro{NOON}{\acl{NO} \acl{ON}}
    \acro{NPE}{non-parallelity error}
    \acro{NSF}{National Science Foundation}
    \acro{OEP}{optimized effective potential}
    \acro{oMCPDFT}[$\om$-MC-PDFT]{range-separated \acl{MC} \acl{1H-PDFT}}
    \acrodef{ON}{occupation number}
    \acro{ORMAS}{occupation-restricted multiple active-space}
    \acro{OTPD}{on-top pair-density}
    \acro{PBE0}{hybrid-\acs{PBE}}
    \acro{PBE}{Perdew-Burke-Ernzerhof}
    \acro{pCCD-lDFT}[pCCD-$\la$DFT]{\acl{pCCD} $\la$\acs{DFT}}
    \acro{pCCD}{pair coupled-cluster doubles}
    \acro{PDFT}{pair-\acl{DFT}}
    \acro{PEC}{potential energy curve}
    \acro{PES}{potential energy surface}
    \acro{PKZB}{Perdew-Kurth-Zupan-Blaha}
    \acro{pp-RPA}{particle-particle \acl{RPA}}
    \acro{PT}{perturbation theory}
    \acro{PT2}{second-order \acl{PT}}
    \acro{PW91}{Perdew-Wang 91}
    \acro{QTAIM}{quantum theory of atoms in molecules}
    \acro{RASSCF}{restricted active-space \acl{SCF}}
    \acro{RDM}{reduced density matrix}
    \acro{revPBE}{revised \acs{PBE}}
    \acro{RL}{Riemann-Liouville}
    \acro{RMSD}{root mean square deviation}
    \acro{RPA}{random-phase approximation}
    \acro{RSH}{range-separated hybrid}
    \acro{SCAN}{strongly constrained and appropriately normed}
    \acro{SCF}{self-consistent field}
    \acro{SDP}{semidefinite programming}
    \acro{SE}{Schr\"odinger equation}
    \acro{SF-CCSD}{\acl{SF}-\acl{CCSD}}
    \acro{SF}{spin-flip}
    \acro{SIE}{self-interaction error}
    \acrodef{SI}{Supporting Information}
    \acro{SNIAD}{spherical \acl{NIAD}}
    \acro{SOGGA11}{second-order \acl{GGA}}
    \acro{SR}{short-range}
    \acro{STO}{Slater-type orbital}
    \acro{SVWN3}{Slater and Vosko-Wilk-Nusair random-phase approximation expression III}
    \acro{tBLYP}{translated \acl{BLYP}}
    \acro{TMAE}[$\MAE$]{total \acl{MAE} per electron}
    \acro{TNIAD}[$\NIAD$]{total normed integral absolute deviation}
    \acro{tPBE}{translated \acl{PBE}}
    \acro{TPSS}{Tao-Perdew-Staroverov-Scuseria}
    \acro{trevPBE}{translated \acs{revPBE}}
    \acro{tr}{conventional translation}
    \acro{tSVWN3}{translated \acl{SVWN3}}
    \acro{TS}{transition state}
    \acro{v2RDM-CASSCF-PDFT}{\acl{v2RDM} \acl{CASSCF} \acl{PDFT}}
    \acro{v2RDM-CASSCF}{\acl{v2RDM}-driven \acl{CASSCF}}
    \acro{v2RDM-CAS}{\acl{v2RDM}-driven \acl{CAS}}
    \acro{v2RDM-DOCI}{\acl{v2RDM}-\acl{DOCI}}
    \acro{v2RDM}{variational \acl{2-RDM}}
    \acro{VB}{valence bond}
    \acro{VO}{variable-order}
    \acro{wB97X}[$\omega$B97X]{$\omega$B97X}
    \acro{WFT}{wave function theory}
    \acro{WF}{wave function}
    \acro{XC}{exchange-correlation}
    \acro{ZPE}{zero-point energy}
    \acro{ZPVE}{zero-point vibrational energy}
\end{acronym}

\newpage

{\bf References}
\vspace{-28pt}

\bibliography{ms}

\end{document}

%---------------------------------------------------
%                Equations skip control
%---------------------------------------------------
%\setlength{\abovedisplayskip}{4pt plus 1.5pt minus 2pt}
%\setlength{\belowdisplayskip}{4pt plus 1.5pt minus 2pt}
%\setlength{\abovedisplayskip}{4pt plus .7pt minus 2pt}
%\setlength{\belowdisplayskip}{4pt plus .7pt minus 2pt}
%---------------------------------------------------

%=================================
% Affiliations
%---------------------------------
\author{Reza Hemmati$^{1,2,\dagger,*}$, Mohammad Mostafanejad$^{1,2,\ddagger,*}$, and J. V. Ortiz$^{3,\star}$ \\[3pt]
    \small $^{1}$\textit{Department of Chemistry, Virginia Tech, Blacksburg, Virginia 24061, USA} \\
    \small $^{2}$\textit{Molecular Sciences Software Institute, Blacksburg, Virginia 24060, USA}\\
    \small $^{3}$\textit{Department of Chemistry and Biochemistry, Auburn University, \\
        Auburn, Alabama, 36849-5312, USA}\\
    \small \textit{E-mails: $^{\dagger}$rhemmati@vt.edu, $^{\ddagger}$smostafanejad@vt.edu, $^{\star}$ortiz@auburn.edu}, \\
    \small $^{*}$Equal contributions \\}
% \author{Reza Hemmati}
% \email{rhemmati@vt.edu}
% \thanks{Equal contribution}
% \author{Mohammad Mostafanejad}
% \email{smostafanejad@vt.edu}
% \thanks{Equal contribution}
% \affiliation{Department of Chemistry, Virginia Tech, Blacksburg, Virginia 24061, USA}
% \affiliation{Molecular Sciences Software Institute, Blacksburg, Virginia 24060, USA}
% %---------------------------------
% \author{Joseph Vincent Ortiz}
% \email{ortiz@auburn.edu}
% \affiliation{Department of Chemistry and Biochemistry, Auburn University, Auburn, Alabama, 36849-5312, USA}
%=================================

%=================================
\title{Supplemental Information: \\ Numerical Analysis of the Complete Active-Space Extended Koopmans's Theorem}
%=================================
\maketitle

%=================================
% Table of Contents
%---------------------------------
\tableofcontents
%=================================
\newpage

%=================================
% Supplemental Info
%---------------------------------
\acresetall	% reset all acronyms
%---------------------------------

%+++++++++++++++++++++++++++++++++
\section{The \acl{EKT} \aclp{IE} of \ce{H2}}\label{SISEC:H2IPS}
%+++++++++++++++++++++++++++++++++
Table \ref{SITAB:H2IPS} presents the first \ac{EKT} \acp{IE} of the \ce{H2}
molecule calculated using the \ac{CASCI} and the \ac{FCI} methods, and the
(aug-)cc-pVXZ basis sets where X = D, T, Q, and 5.
\begin{table*}[!htbp]
    \centering
    \setlength{\tabcolsep}{3pt}
    \setlength{\extrarowheight}{1pt}
    \caption{The first \acs{EKT} ground state ionization energy of
        \ce{H2} molecule calculated with \acs{CASCI} and \acs{FCI} methods and
        (aug-)cc-pVXZ basis sets where X = D, T, Q, and 5$^{a}$}
    \label{SITAB:H2IPS}
    % \resizebox{\textwidth}{!}{
    \begin{tabular}{lccccc}
        \hline\hline
        Level of Theory                 & X & $\Etot$   & $\IEKTm$ & $\EKTm$  & $\tau_m$$^b$ \\
        \hline
        \mr{4}{*}{CAS(2,4)/cc-pVXZ}     & D & -1.152976 & 0.588941 & 0.588941 & 0.000000     \\
                                        & T & -1.163913 & 0.598314 & 0.598314 & 0.000000     \\
                                        & Q & -1.164719 & 0.598396 & 0.598396 & 0.000000     \\
                                        & 5 & -1.163005 & 0.621804 & 0.621804 & 0.000000     \\[5pt]
        \mr{4}{*}{CAS(2,4)/aug-cc-pVXZ} & D & -1.157337 & 0.619167 & 0.619167 & 0.000000     \\
                                        & T & -1.162273 & 0.621458 & 0.621458 & 0.000000     \\
                                        & Q & -1.162846 & 0.621754 & 0.621754 & 0.000000     \\
                                        & 5 & -1.163008 & 0.621809 & 0.621809 & 0.000000     \\
        \hline\hline
    \end{tabular}
    % }
    \begin{tablenotes}
        \item \qquad \qquad \quad $^a$ All energy values are in atomic units.
        \item \qquad \qquad \quad $^b$ The $m$ is either \acs{CASCI} or
        \acs{FCI}.
    \end{tablenotes}
\end{table*}
The results of Table \ref{SITAB:H2IPS}, while conformant with the inequality 9
in the main text, exhibit two types of irregularities which violate the
variation principle.\cite{Helgaker:2000:BOOK,Szabo:1996:BOOK} First, the total
energy of the \ce{H2} calculated at the CAS(2,4)/cc-pV5Z level of theory is
higher than those of the CAS(2,4)/cc-pVXZ calculations where X = T, and Q.
Second, the total energies calculated at the CAS(2,4)/cc-pVXZ level of theory
are lower than those of the CAS(2,4)/aug-cc-pVXZ in which X = T and Q. We have
tested various optimizers and initial orbitals for the \ac{CASCI} calculations
to ensure that the irregularities are not due to the convergence issues. As
such, the reason behind these irregularities is not clear to us.
%+++++++++++++++++++++++++++++++++
\section{Normalization of the \acl{EKT} \acl{IE} surface data}\label{SISEC:STANDARD}
%+++++++++++++++++++++++++++++++++
We use the \textsc{MinMaxScaler} class of the \sklearn\
package\cite{Pedregosa:2011:2825} to normalize the \ac{EKT} \ac{IE} surface data
within the range $[0,1]$, as shown in Figs.~\ref{SIFIG:IEKT3D} and
\ref{SIFIG:TAU3D} of \si\ as well as Fig.~2 of the main text. The Min-Max
transformation can be expressed as

\begin{gather}\label{EQ:MINMAX}
    I_{\text{std}} = \frac{I_{i} - I_{\text{min}}}{I_{\text{max}} - I_{\text{min}}}    \\
    I_{\text{norm}} = I_{\text{std}} \times (U - L) + L
\end{gather}
where $U$ and $L$ are the upper and lower limits of the target range which are
set to one and zero, respectively. The symbols, $I_{\text{std}}$ and
$I_{\text{norm}}$ denote the standardized and normalized \acp{IE}, respectively.
Also, $I_{i}$, $I_{\text{min}}$, and $I_{\text{max}}$ represent the $i$th,
minimum and maximum values \acp{IE} in the dataset, respectively. We have
summarized the transformation parameters in Table~\ref{SITAB:STATS}.

\begin{table*}[!htbp]
    \centering
    \setlength{\tabcolsep}{3pt}
    \setlength{\extrarowheight}{1pt}
    \caption{The transformation parameters required for the normalization of the \acl{IE}
        surface data}
    \label{SITAB:STATS}
    % \resizebox{\textwidth}{!}{
    \begin{tabular}{lccccc}
        \hline\hline
        Bond Length ($R$)$^{a}$ & $\min(\IEKTCAS)$$^b$ & $\max(\IEKTCAS)$$^b$ & Range (D)$^b$ & Scale ($\al$)  & Shift ($\be$)   \\
                \hline
        $\Req$                  & 0.278464             & 0.278467             & 2.6e-06       & 381781.391974  & -106312.532623  \\
        $5\Req$                 & 0.196311             & 0.196312             & 1.4e-07       & 7183908.046034 & -1410283.723431 \\
        $15\Req$                & 0.196311             & 0.196311             & 1.2e-07       & 8503401.361408 & -1669309.580952 \\
        \hline\hline
    \end{tabular}
    % }
    \begin{tablenotes}
        \item \quad $^a$ All bond lengths are in \AA.
        \item \quad $^b$ All energies are in atomic units.
    \end{tablenotes}
\end{table*}

The scale, $\al$, and shift, $\be$, parameters can be calculated using the following
equations:

\begin{gather}
    \al = \frac{U - L}{I_{\text{max}} - I_{\text{min}}}, \\
    \be = L - \al \times I_{\text{min}}.
\end{gather}
%=================================

%+++++++++++++++++++++++++++++++++
\section{Extended \acl{KT} ionization energy surface of \ce{LiH}}\label{SISEC:IP3D}
%+++++++++++++++++++++++++++++++++
In order to investigate the Ernzerhof's conjecture,\cite{Ernzerhof:2009:793} we
have generated the three-dimensional \ac{CASCI} \ac{EKT} ionization energy
surface of \ce{LiH} molecule in a multi-step process. After setting up the
\ce{LiH} molecule by fixing the \ce{H} atom at the origin and placing the
\ce{Li} atom on the $z$-axis at $z = \Req = 1.5949$ \AA, the molecule is
embedded in a two-dimensional mesh of grid points in the $yz$ plane. The
boundaries of our rectangular mesh range from $y\in [-2.0,2.0)$ and $z\in
[-2.0,20.0)$ with the step-sizes of 0.1 and 0.2 \AA, respectively. We scan the
bond length across five separate internuclear distances of $c \Req$, where $c =$
1, 3, 5, 10 and 15. The combination of the bond lengths and the grid points
results in a total of 22,000 single-point \ac{HF} and subsequent \ac{CASCI}
calculations on both $N$- and ($N\!-\!1$)-electron states of \ce{LiH}. At each
bond length, $R$, the individual $(y,z)$ grid points are probed by moving a
single diffuse primitive $s$-type Gaussian basis function with a fixed exponent
$\al=0.0076000$ in the $yz$ plane. By symmetry, the diffuse primitive function
is allowed to interact with the active orbitals of the same symmetry in the
\ce{LiH} molecule. For each $(R,y,z)$ point, we perform a
\ac{HF}/aug-cc-pwCVTZ\cite{Dunning:1989:1007,Peterson:2002:10548,
Kendall:1992:6796} calculation on the neutral (singlet) ground state
wavefunction of \ce{LiH} and use the resulting orbitals for the subsequent
CAS(2,2)/aug-cc-pwCVTZ computation on the same electronic state. The resulting
\ac{CASCI} orbitals are subsequently used for the CAS(1,2)/aug-cc-pwCVTZ energy
calculation, this time on the doublet cationic state of \ce{LiH}.

\begin{figure}[!tbph]
    \centering
    \setlength{\abovecaptionskip}{-2pt}
    \includegraphics[scale=0.22]{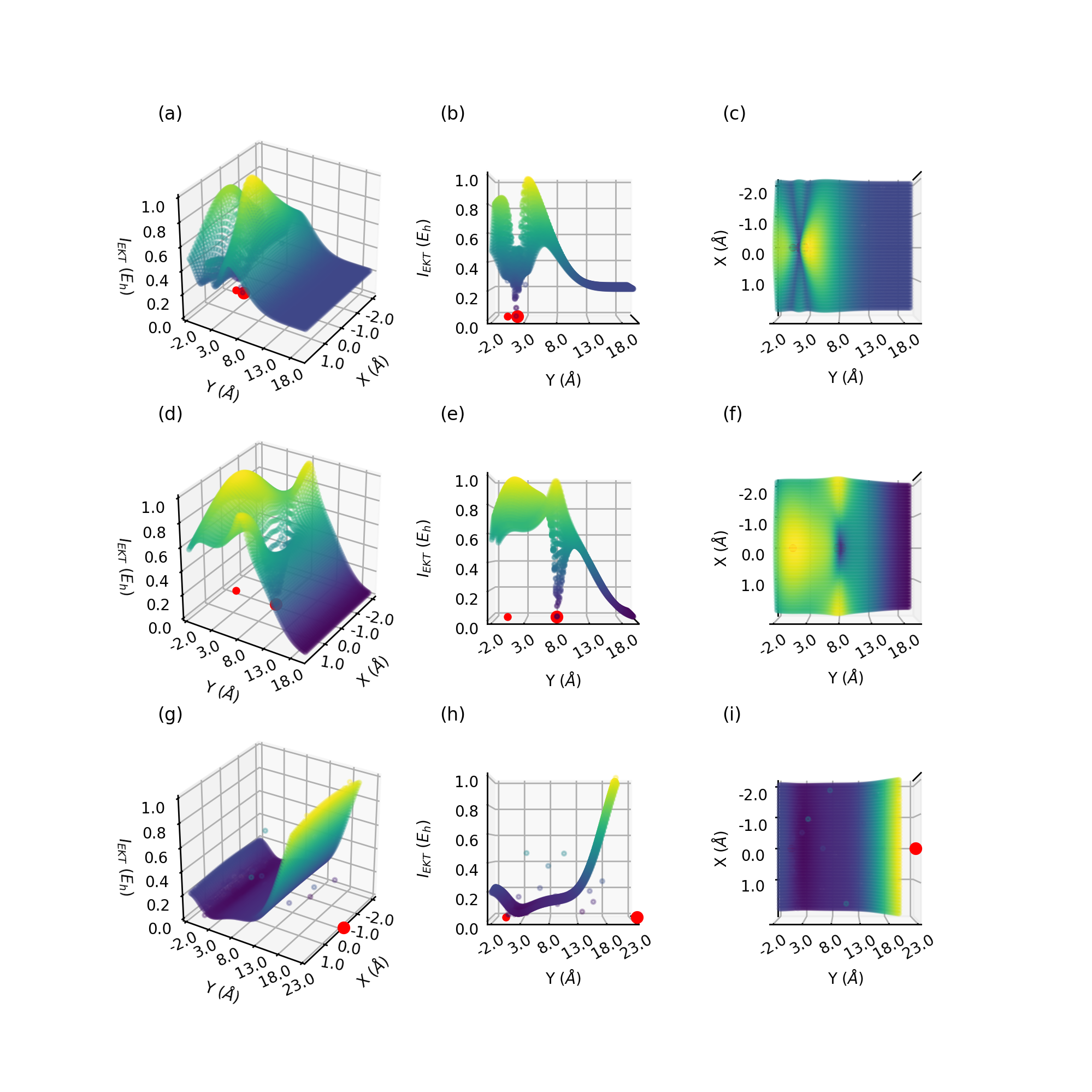}
    \caption{The first \ac{EKT} ionization energy surface of \ce{LiH}
        molecule calculated at the CAS(2,2)/aug-cc-pwCVTZ level of theory. The bond
        lengths are fixed at (a--c) $R=\Req$, (d--f) $R=5\Req$, and (g--i)
        $R=15\Req$ where $\Req=1.5949$ \AA.}
    \label{SIFIG:IEKT3D}
\end{figure}

\begin{figure}[!tbph]
    \centering
    \setlength{\abovecaptionskip}{-2pt}
    \includegraphics[scale=0.2]{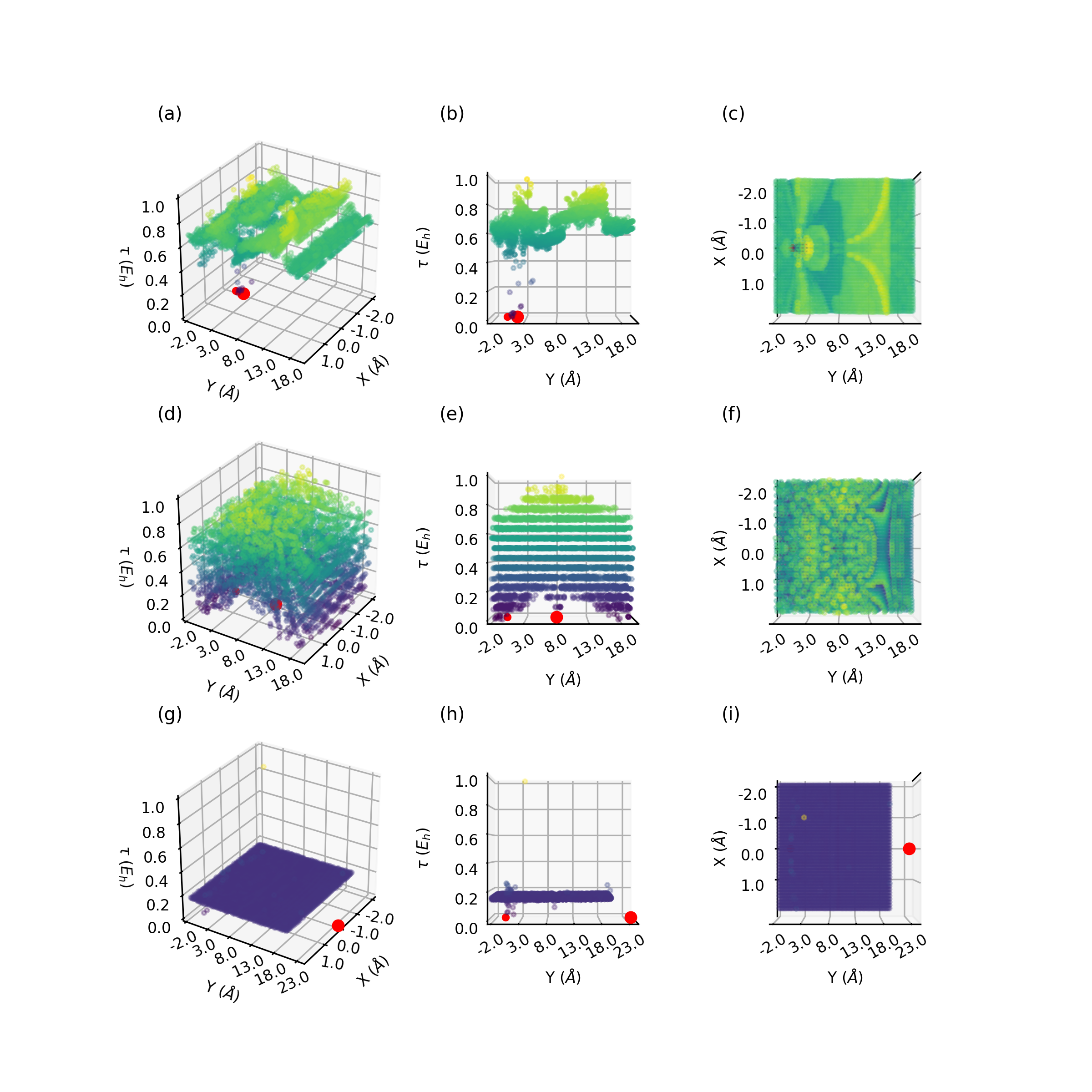}
    \caption{Numerical error, $\TAUCAS$, in the first \ac{EKT} ionization energy
        surface of \ce{LiH} molecule calculated at the CAS(2,2)/aug-cc-pwCVTZ
        level of theory. The bond lengths are fixed at (a--c) $R=\Req$, (d--f)
        $R=5\Req$, and (g--i) $R=15\Req$ where $\Req=1.5949$ \AA.}
    \label{SIFIG:TAU3D}
\end{figure}

Figure \ref{SIFIG:IEKT3D} shows that at the equilibrium bond length, the
\ac{EKT} ionization energy values become smaller around the middle bond and
closer to the \ce{Li} atom than those in other regions of space. The \ac{EKT}
\ac{IE} surface also shows multiple minima and two individual maxima around the
\ce{H} and \ce{Li} atoms the latter being more pronounced. As one moves away
from the molecule along the $z$-axis, the \ac{EKT} ionization energy values
rapidly decay to a constant value which is consistent with our previous
observations for the $\TAUCAS$ values in Tables~I--V of the main text. At the
stretched bond length of $R=5\Req$, the \ac{EKT} \ac{IE} surface exhibits a
distinct minimum around the \ce{Li} atom and a maximum around the \ce{H} atom
alongside two maxima on the two sides of the \ce{Li} atom. At the bond length
$R=15\Req$, the \ac{EKT} \ac{IE} surface shows increases in the energy values
around the \ce{Li} atom and to a lesser extent around the \ce{H} atom. It is
worth noting that the \ac{EKT} \ac{IE} surface values are changing in the range
of $10^{-6}-10^{-7} \Eh$ across the aforementioned bond lengths which is
significant for highly accurate spectroscopic studies but may not be important
for most chemical applications. Figure \ref{SIFIG:TAU3D} shows the numerical
error, $\TAUCAS$, in the \ac{CASCI} \ac{EKT} ionization energy surface of
\ce{LiH} molecule shown in Fig.~\ref{SIFIG:IEKT3D}.
At the equilibrium bond length, the numerical error, $\TAUCAS$, in the \ac{EKT}
\ac{IE} surface is smaller around the nuclear positions and at other positions
away from the nuclei, changes roughly within a constant range of a microhartree.
As the bond length increases, the magnitude of the numerical error decreases by
about an order of magnitude which is also lower than the energy convergence
threshold for our \ac{CASCI} calculations ($1\ \mu \Eh$).

%\section{Acronyms}
\begin{acronym}
    \acrodefplural{1-RDM}{one-electron \aclp{RDM}}
    \acrodefplural{2-RDM}{two-electron \aclp{RDM}}
    \acrodefplural{3-RDM}{three-electron \aclp{RDM}}
    \acrodefplural{4-RDM}{four-electron \aclp{RDM}}
    \acrodefplural{EA}{electron affinities}
    \acrodefplural{IE}{ionization energies}
    \acrodefplural{RDM}{reduced density matrices}
    \acro{1-HRDM}{one-hole \acl{RDM}}
    \acro{1-RDM}{one-electron \acl{RDM}}
    \acrodef{1H-PDFT}{one-parameter hybrid \acl{PDFT}}
    \acro{2-HRDM}{two-hole \acl{RDM}}
    \acro{2-RDM}{two-electron \acl{RDM}}
    \acro{3-RDM}{three-electron \acl{RDM}}
    \acro{4-RDM}{four-electron \acl{RDM}}
    \acro{ACI}{adaptive \acl{CI}}
    \acro{ACI-DSRG-MRPT2}{\acl{ACI}-\acl{DSRG} \acl{MR} \acl{PT2}}
    \acro{ACSE}{anti-Hermitian \acl{CSE}}
    \acrodef{AKEEPE}[$\AKEEe$]{absolute kinetic energy error per electron}
    \acrodef{AKEE}{absolute kinetic energy error}
    \acro{AO}{atomic orbital}
    \acro{AQCC}{averaged quadratic \acl{CC}}
    \acro{ARC}{Advanced Research Computing}
    \acro{aug-cc-pVQZ}{augmented correlation-consistent polarized-valence quadruple-$\ze$}
    \acro{aug-cc-pVTZ}{augmented correlation-consistent polarized-valence triple-$\ze$}
    \acro{aug-cc-pwCV5Z}[aug-cc-pwCV5Z]{augmented correlation-consistent polarized weighted core-valence quintuple-$\ze$}
    \acro{B3LYP}{Becke-3-\acl{LYP}}
    \acro{BLA}{bond length alternation}
    \acro{BLYP}{Becke and \acl{LYP}}
    \acro{BO}{Born-Oppenheimer}
    \acro{BP86}{Becke 88 exchange and P86 Perdew-Wang correlation}
    \acro{BPSDP}{boundary-point \acl{SDP}}
    \acro{CAM}{Coulomb-attenuating method}
    \acro{CAM-B3LYP}{Coulomb-attenuating method \acl{B3LYP}}
    \acro{CASCI}[CAS-CI]{\acl{CAS} \acl{CI}}
    \acro{CAS-PDFT}{\acl{CAS} \acl{PDFT}}
    \acro{CASPT2}{\acl{CAS} \acl{PT2}}
    \acro{CASSCF}{\acl{CAS} \acl{SCF}}
    \acro{CAS}{complete active-space}
    \acro{cc-pVDZ}{correlation-consistent polarized-valence double-$\ze$}
    \acro{cc-pVTZ}{correlation-consistent polarized-valence triple-$\ze$}
    \acro{CCSDT}{coupled-cluster, singles doubles and triples}
    \acro{CCSD}{coupled-cluster with singles and doubles}
    \acro{CC}{coupled-cluster}
    \acro{CI}{configuration interaction}
    \acro{CO}{constant-order}
    \acro{CPO}{correlated participating orbitals}
    \acro{CSF}{configuration state function}
    \acro{CS-KSDFT}{\acl{CS}-\acl{KSDFT}}
    \acro{CSE}{contracted \acl{SE}}
    \acro{CS}{constrained search}
    \acro{DC-DFT}{density corrected-\acl{DFT}}
    \acro{DC-KDFT}{density corrected-\acl{KDFT}}
    \acro{DE}{delocalization error}
    \acro{DFT}{density functional theory}
    \acro{DF}{density-fitting}
    \acro{DIIS}{direct inversion in the iterative subspace}
    \acro{DMRG}{density matrix renormalization group}
    \acro{DOCI}{doubly occupied \acl{CI}}
    \acro{DSRG}{driven similarity renormalization group}
    \acro{EA}{electron affinity}
    \acro{EKT}{extended Koopmans's theorem}
    \acro{ERI}{electron-repulsion integral}
    \acro{EUE}{effectively unpaired electron}
    \acro{FC}{fractional calculus}
    \acro{FCI}{full \acl{CI}}
    \acro{FP-1}{frontier partition with one set of interspace excitations}
    \acro{FSE}{fractional \acl{SE}}
    \acro{ftBLYP}{fully \acl{tBLYP}}
    \acro{ftPBE}{fully \acl{tPBE}}
    \acro{ftSVWN3}{fully \acl{tSVWN3}}
    \acro{ft}{full translation}
    \acro{GASSCF}{generalized active-space \acl{SCF}}
    \acro{GGA}{generalized gradient approximation}
    \acro{GL}{Gr\"unwald-Letnikov}
    \acro{GMCPDFT}[G-MC-PDFT]{generalized \acl{MCPDFT}}
    \acro{GTO}{Gaussian-type orbital}
    \acro{HCI}{hole-state \acl{CI}}
    \acro{HF}{Hartree-Fock}
    \acro{HISS}{Henderson-Izmaylov-Scuseria-Savin}
    \acrodef{HK}{Hohenberg-Kohn}
    \acro{HOMO}{highest-occupied \acl{MO}}
    \acro{HONO}{highest-occupied \acl{NO}}
    \acro{HPC}{high-performance computing}
    \acro{HPDFT}{hybrid \acl{PDFT}}
    \acro{HRDM}{hole \acl{RDM}}
    \acro{HSE}{Heyd-Scuseria-Ernzerhof}
    \acro{HXC}{Hartree-\acl{XC}}
    \acro{IE}{ionization energy}
    \acro{IPEA}{ionization potential electron affinity}
    \acro{IPSDP}{interior-point \acl{SDP}}
    \acro{KDFT}{kinetic \acl{DFT}}
    \acro{KSDFT}[KS-DFT]{\acl{KS} \acl{DFT}}
    \acro{KS}{Kohn-Sham}
    \acro{KT}{Koopmans's theorem}
    \acro{LBFGS}[L-BFGS]{limited-memory Broyden-Fletcher-Goldfarb-Shanno}
    \acro{LC}{long-range corrected}
    \acro{LC-VV10}{\acl{LC} Vydrov-van Voorhis 10}
    \acro{l-DFVB}[$\la$-DFVB]{$\la$-density functional \acl{VB}}
    \acro{LEB}{local energy balance}
    \acro{LE}{localization error}
    \acrodef{lftBLYP}[$\la$-\acs{ftBLYP}]{$\la$-\acl{ftBLYP}}
    \acrodef{lftPBE}[$\la$-\acs{ftPBE}]{$\la$-\acl{ftPBE}}
    \acrodef{lftrevPBE}[$\la$-\acs{ftrevPBE}]{$\la$-\acl{ftrevPBE}}
    \acrodef{lftSVWN3}[$\la$-\acs{ftSVWN3}]{$\la$-\acl{ftSVWN3}}
    \acro{lMCPDFT}[$\la$-MC-PDFT]{\acl{MC} \acl{1H-PDFT}}
    \acro{LMF}{local mixing function}
    \acro{LO}{Lieb-Oxford}
    \acro{LP}{linear programming}
    \acro{LR}{long-range}
    \acro{LSDA}{local spin-density approximation}
    \acrodef{ltBLYP}[$\la$-\acs{tBLYP}]{$\la$-\acl{tBLYP}}
    \acrodef{ltPBE}[$\la$-\acs{tPBE}]{$\la$-\acl{tPBE}}
    \acrodef{ltrevPBE}[$\la$-\acs{trevPBE}]{$\la$-\acl{trevPBE}}
    \acrodef{ltSVWN3}[$\la$-\acs{tSVWN3}]{$\la$-\acl{tSVWN3}}
    \acro{LUMO}{lowest-unoccupied \acl{MO}}
    \acro{LUNO}{lowest-unoccupied \acl{NO}}
    \acro{LYP}{Lee-Yang-Parr}
    \acro{M06}{Minnesota 06}
    \acro{M06-2X}{\acl{M06} with double non-local exchange}
    \acro{M06-L}{\acl{M06} local}
    \acro{MAEPE}[$\MAEe$]{\acl{MAE} per electron}
    \acro{MAE}{mean absolute error}
    \acro{MAKEE}[$\MAKEE$]{mean absolute kinetic energy error per electron}
    \acro{MAX}{maximum absolute error}
    \acro{MC1H-PDFT}{\acl{MC} \acl{1H-PDFT}}
    \acro{MC1H}{\acl{MC} one-parameter hybrid \acl{PDFT}}
    \acro{MCHPDFT}{\acl{MC} hybrid-\acl{PDFT}}
    \acro{MCPDFT}[MC-PDFT]{\acl{MC} \acl{PDFT}}
    \acro{MCRSHPDFT}[$\mu\la$-MCPDFT]{\acl{MC} range-separated hybrid-\acl{PDFT}}
    \acro{MCSCF}{\acl{MC} \acl{SCF}}
    \acrodef{MC}{multiconfiguration}
    \acro{MN15}{Minnesota 15}
    \acro{MOLSSI}[MolSSI]{Molecular Sciences Software Institute}
    \acro{MO}{molecular orbital}
    \acro{MP2}{second-order M\o ller-Plesset \acl{PT}}
    \acro{MR-AQCC}{\acl{MR}-averaged quadratic \acl{CC}}
    \acrodef{MR}{multireference}
    \acro{MS0}{MS0 meta-GGA exchange and revTPSS GGA correlation}
    \acro{NDDO}{neglect of diatomic differential overlap}
    \acro{NGA}{nonseparable gradient approximation}
    \acro{NIAD}{normed integral absolute deviation}
    \acro{NO}{natural orbital}
    \acro{NOON}{\acl{NO} \acl{ON}}
    \acro{NPE}{non-parallelity error}
    \acro{NSF}{National Science Foundation}
    \acro{OEP}{optimized effective potential}
    \acro{oMCPDFT}[$\om$-MC-PDFT]{range-separated \acl{MC} \acl{1H-PDFT}}
    \acrodef{ON}{occupation number}
    \acro{ORMAS}{occupation-restricted multiple active-space}
    \acro{OTPD}{on-top pair-density}
    \acro{PBE0}{hybrid-\acs{PBE}}
    \acro{PBE}{Perdew-Burke-Ernzerhof}
    \acro{pCCD-lDFT}[pCCD-$\la$DFT]{\acl{pCCD} $\la$\acs{DFT}}
    \acro{pCCD}{pair coupled-cluster doubles}
    \acro{PDFT}{pair-\acl{DFT}}
    \acro{PEC}{potential energy curve}
    \acro{PES}{potential energy surface}
    \acro{PKZB}{Perdew-Kurth-Zupan-Blaha}
    \acro{pp-RPA}{particle-particle \acl{RPA}}
    \acro{PT}{perturbation theory}
    \acro{PT2}{second-order \acl{PT}}
    \acro{PW91}{Perdew-Wang 91}
    \acro{QTAIM}{quantum theory of atoms in molecules}
    \acro{RASSCF}{restricted active-space \acl{SCF}}
    \acro{RDM}{reduced density matrix}
    \acro{revPBE}{revised \acs{PBE}}
    \acro{RL}{Riemann-Liouville}
    \acro{RMSD}{root mean square deviation}
    \acro{RPA}{random-phase approximation}
    \acro{RSH}{range-separated hybrid}
    \acro{SCAN}{strongly constrained and appropriately normed}
    \acro{SCF}{self-consistent field}
    \acro{SDP}{semidefinite programming}
    \acro{SE}{Schr\"odinger equation}
    \acro{SF-CCSD}{\acl{SF}-\acl{CCSD}}
    \acro{SF}{spin-flip}
    \acro{SIE}{self-interaction error}
    \acrodef{SI}{Supporting Information}
    \acro{SNIAD}{spherical \acl{NIAD}}
    \acro{SOGGA11}{second-order \acl{GGA}}
    \acro{SR}{short-range}
    \acro{STO}{Slater-type orbital}
    \acro{SVWN3}{Slater and Vosko-Wilk-Nusair random-phase approximation expression III}
    \acro{tBLYP}{translated \acl{BLYP}}
    \acro{TMAE}[$\MAE$]{total \acl{MAE} per electron}
    \acro{TNIAD}[$\NIAD$]{total normed integral absolute deviation}
    \acro{tPBE}{translated \acl{PBE}}
    \acro{TPSS}{Tao-Perdew-Staroverov-Scuseria}
    \acro{trevPBE}{translated \acs{revPBE}}
    \acro{tr}{conventional translation}
    \acro{tSVWN3}{translated \acl{SVWN3}}
    \acro{TS}{transition state}
    \acro{v2RDM-CASSCF-PDFT}{\acl{v2RDM} \acl{CASSCF} \acl{PDFT}}
    \acro{v2RDM-CASSCF}{\acl{v2RDM}-driven \acl{CASSCF}}
    \acro{v2RDM-CAS}{\acl{v2RDM}-driven \acl{CAS}}
    \acro{v2RDM-DOCI}{\acl{v2RDM}-\acl{DOCI}}
    \acro{v2RDM}{variational \acl{2-RDM}}
    \acro{VB}{valence bond}
    \acro{VO}{variable-order}
    \acro{wB97X}[$\omega$B97X]{$\omega$B97X}
    \acro{WFT}{wave function theory}
    \acro{WF}{wave function}
    \acro{XC}{exchange-correlation}
    \acro{ZPE}{zero-point energy}
    \acro{ZPVE}{zero-point vibrational energy}
\end{acronym}

\newpage

{\bf References}
\vspace{-28pt}

\bibliography{supplement}